\documentclass{article}
\usepackage{enumerate,amsmath,amsthm,amssymb,latexsym,
amsfonts,amscd}
\usepackage{xypic}
\xyoption{curve}
\usepackage[mathscr]{eucal}
\usepackage[all]{xy}
\usepackage{graphicx}
\usepackage{float}
\theoremstyle{plain}

\theoremstyle{definition}

\newtheorem{example}{Example}[section]
\newtheorem{remark}{Remark}[section]

\numberwithin{equation}{section}

\newcommand{\F}{\mathcal F}

\newcommand{\R}{\mathcal R}

\newcommand{\bF}{\mathbb{F}}

\newcommand{\bR}{\mathbb{R}}
\newcommand{\bT}{\mathbb{T}}

\newcommand{\As}{\mathscr A}
\newcommand{\Bs}{\mathscr B}

\newcommand{\grp}{{\mathsf{G}}}

\newcommand{\obj}{{\rm Obj}}

\newcommand{\what}{\widehat}

\newcommand{\ul}{\underline}

\renewcommand{\a}{\alpha}
\newcommand{\be}{\beta}

\newcommand{\del}{\partial}

\newcommand{\med}{\medbreak}
\newcommand{\medn}{\medbreak\noindent}

\newcommand{\lra}{{\longrightarrow}}

\newcommand{\ovsetLR}[1]{\overset {#1}{\Longrightarrow}}
\newcommand{\ovsetLL}[1]{\overset {#1}{\Longleftarrow}}

\begin{document}

\title{\textbf{Rate distortion coevolutionary dynamics\\
and the flow nature of cognitive epigenetic systems}}

\author{James F. Glazebrook\footnote{Department of Mathematics and
Computer Science, Eastern Illinois University, 600 Lincoln Avenue,
Charleston IL 61920--3099 USA, email jfglazebrook@eiu.edu}, PhD\\
Department of Mathematics and Computer Science\\ Eastern Illinois
University\\and\\Rodrick Wallace\footnote{549 W. 123 St., Suite
16F, New York, NY, 10027, USA. Telephone (212) 865-4766, email
rdwall@ix.netcom.com. Affiliation is for identification only.},
PhD\\Division of Epidemiology\\ The New York State Psychiatric
Institute}

\maketitle

\begin{abstract}
We outline a model for a cognitive epigenetic system based on
elements of the Shannon theory of information and the statistical physics of
the generalized Onsager relations.
Particular attention is paid to the concept of the rate distortion function
and from another direction as motivated by the thermodynamics of computing,
the fundamental homology with the free energy density of a physical system.
A unifying aspect of the dynamic framework involves the concept of a groupoid
and of a groupoid atlas. From a stochastic differential equation we postulate a
multidimensional It\^{o} process for an epigenetic system from which a stochastic flow
may permeate through components of this atlas.
\end{abstract}


\section{Introduction}

Living systems as far-from-equilibrium open systems, are essentially cognitive, and conversely,
in order to grasp the essence of cognition, one may attempt to understand the ontology of the former processes.
Often this is viewed in the framework of a cell-like, structured, self-organizing system engaged in a two-way interaction between
its functional mechanisms and that of its neighboring environment. There are several theories that
revolve around this principle. Readers may be familiar with a central relationship as explained by
autopoiesis, a seemingly enduring theory as developed by Maturana and Varela (1980a, 1980b)
over several decades. Though a much earlier and somewhat different hypothesis of Bertalanffy (1972)
had proposed that living systems are maintained in non-equilibrium states by a flow-like patterns
 when drawing matter and energy from their environment, and adjust accordingly in a ``flowing balance''
 (Bertalanffy, 1972; Capra, 1996).
 Further scientific rigor aimed at understanding this hypothesis was achieved by
Prigogine (1980) who formulated similar ideas cast within a
theory of dissipative structures.
Whereas living systems are continuously maintained in-far-from equilibrium, dissipative
structures do likewise while being capable of evolving. They are destabilized in the increase of information and
energy, though as they remain self-organized and self-perpetuating, the complexity of their structure increases; often
this is simply for the sake of survival within the environment. Elucidating a possible synthesis of these separate approaches, commencing from
the Bertalanffy hypothesis, is
the main topic of Capra (1996) . But cognition is also a function
of a prevailing culture: by means of social and historical patterns of behavior traditions, etc., human cognition feeds back into that culture
through the course of social interaction, adaptive technologies, policies, memetic trends, etc., and inevitably alters
it in time (see e.g. Clark, 1997; Hollan et al., 2000; Hutchins, 1994; Richerson and Boyd, 2004; Wallace and Fullilove, 2008).

There are several common factors at stake here, and in seeking to understand these,
the `immunology--language' viewpoint of Atlan and Cohen (1998) (see also Cohen, 2000) views human organizations at all levels as
perceiving patterns of threat or opportunity, comparing those patterns with
some internal, learned or inherited, picture of the world, and
then choosing one or a small number of responses from a vastly
larger repertory of that which is possible to them. Putting it another way,
consider a basic observation concerning the immune system: the latter
incorporates its own system of options and responds cognitively at the level of information processing
such that the \textit{meaning} of an antigen is defined by the response of the
immune system, somewhat reminiscent of earlier work of Jerne (1974)
who had postulated such `meaning'.
The broader cognitive model that results from this can said to be a `reactive' system
determined by contextual factors in Cohen and Harel (2007) as depicted in Figure 1.

Starting from this basic perspective, Wallace (2005)
has formulated a model of cognition that takes `meaning' in the sense
of Dretske's theory of semantic communication (Dretske, 1981, 1988) claiming that
the immune perception/response information networks of any cognitive system must
be constrained by Shannon's fundamental limit
theorems of information theory (see e.g. Ash, 1990; Berger, 1971; Cover and Thomas, 1991).
These networks comprising of cognitive modules
interact within the framework of a kind of broadcasting
system relaying within `a theater of consciousness' --
the basic operative hypothesis of the \emph{Global (Neuronal) Workspace} theory
 as developed in (Baars, 1988; Baars and Franklin, 2003) that forms an integral part of the model
 in (Wallace, 2005; Wallace and Fullilove, 2008).

Here we take a further step forward that incorporates a number of related factors mainly following
the (generalized) Onsager relations of non-equilibrium thermodynamics
combined with the principles of \emph{rate distortion} theory which is one of the mainstays of Shannon's
pioneering work.
One particular observation is that distortion in communication between interacting cognitive modules,
is patently stochastic, in particular, it is manifestly a process of Brownian motion.
This observation is developed
in several steps relative to a corresponding stochastic differential equation that eventually unfolds to a
kind of master equation. By considering critical solutions for the distortion, we argue on a case-by-case basis
that the ensuing Brownian motion (viz Weiner process) can be locally one of bounded variation. Accordingly, we are motivated
to introduce the necessary conditions that permit integration of these equations in order to obtain \emph{a multidimensional
It\^{o} process}; further we discuss those formal conditions that describe an associated stochastic solution flow.
Thus in a way we have recovered Bertalanffy's perception of the ``flow'' of living systems, the kind of which
information-tied epigenetic systems are certain representations.

Whereas such model conditions are mathematically stated for a flow in a smooth manifold geometry, we do not claim such
`smoothness' in general. Indeed, such epigenetic `flow systems' in living organisms and brain-environment
interactions may be at best continuous over time and are most likely to be `singular' in some sense. The latter requires harder
topological techniques and possible complications that detract from a more natural interpretation. Eventually one seeks a
workable common ground between various mathematical abstractions and the actual empirical properties of the systems themselves.

%%%%%%%%%%%%%%%%%%%%%%%%%%%%%%%%%%%%%%%%%%%%%%%%%%%%%%%%%%%%%%%%%%%%%%%%%%%%%%%%%%%%%%%%%%%%%%%%%%%%%%%

\section{The basis of the epigenetic model}

\subsection{Gene-environment interaction}

Another slant emerging from this cognitive paradigm goes towards epigenetic information sources as providing
a \emph{tunable catalyst} directed to gene expression by which the embedding
of information sources can direct developmental pathways within the ontology of enveloped structures
via the process of information (Wallace and Wallace, 2008). This leads to further developments
of several cognitive paradigms for gene expression, in part furthering the scientific reason
underlying that of Jablonka and Lamb (2005) who consider epigenetic
inheritance systems in which information may be transmitted through generations, not just simply through the base sequence of
DNA, but also transmitted via cultural and behavioral means in higher animals, and by epigenetic means in cell
lineages. This is initiated by memory systems that enable somatic cells of differing phenotype, but of identical genotype,
to transmit their phenotypes to their descendants, even in the absence of the original stimuli that
had engaged these phenotypes.

Conditional upon an individual's phenotype, environmental factors may trigger alterations in behavior and health,
eventually impinging upon the nervous system with the likely consequence of mental disarray. In this respect
Moffitt et al. (2006) (cf Caspi and Moffitt, 2006) make several observations:
(1) a heritable versus environmental influence on phenotype variation across
a given environment, (2) altered gene expressions via epigenetic programming geared in response to subsequent health-behavioral reactions
towards the environment, (3) how an individual's phenotype determines a risk-level towards the environment, and (4) behavioral effects due to
interdependence between specified variations in the DNA sequence specific to a measured environment. Indispensable to understanding
and extending these findings is the parallel question of how gene-cultural interaction, two distinct but interacting
 hereditary systems, compares psychopathology across oriental
and western cultures in basic perceptual processes, often an important piece missing from the larger picture (Nisbett, 2003; Richerson and Boyd, 2004; Wallace, 2005; Wallace, 2009). As such distinct interacting systems of information influencing action and behavior,
both kinds (genes and culture) are claimed in Durham (1991) to create a real and unambiguous symmetry: between genes and phenotypes on the one hand,
and culture and phenotypes on the other, whereby genes and culture may be reasonably viewed as two parallel lines of hereditary influence on phenotypes. From the perspective of (Wallace, 2005; Wallace and Wallace, 2008; Wallace and Wallace, 2009) both can be realized as generalized languages in the sense they
have their own intrinsic recognizable grammar and syntax (see \S\ref{meaningful}). Likewise, on reflecting upon the fundamental mechanisms
of serial endosymbiotic theory (see e.g. Margulis, 2004), Witzany (2006) argues a case for extending the latter via biosemiotic
cell-cell type interactions as `signed' language-communication processes subject to a range of syntactic, pragmatic and semantic rules as
applied to protein coding DNA, RNA editing, DNA splicing, transcription and other essential functions.

\begin{figure}
\begin{center}
\includegraphics[width=80mm]{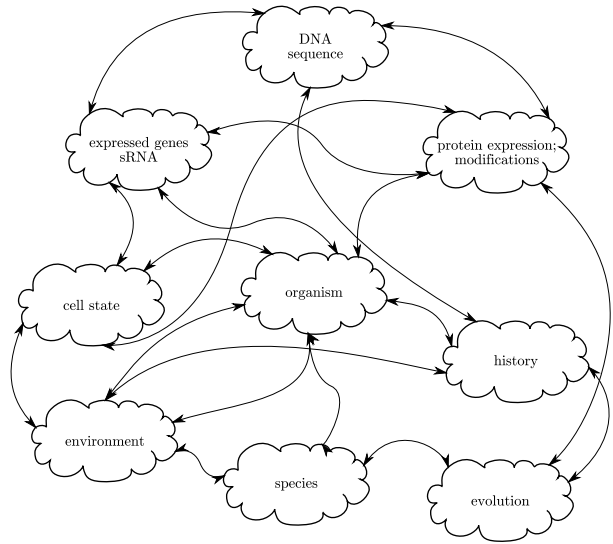}

\caption{An (inter) `reactive' system (as adapted from Fig. 2 p.177 of Cohen and Harel, 2007).
We may interpret this as a large-scale scheme of interacting and reciprocating processes
modeled upon a kind of `atlas' (see below).}
\end{center}
\end{figure}

\subsection{Autopoiesis and the structuralist approach}

A more general perspective is to consider those central relations characterizing the system and its various structural
components, and how a state is altered under perturbations by the environment.
An autopoietic system is organizationally closed and structurally determined. The
system's autopoiesis is preserved within the living state, adaptable only to structural fluctuations for as long
as the living entity survives within and is structurally coupled to its environment; otherwise there is termination.
The autopoiesis of the nervous system (though itself is not strictly autopoietic) functionally self-(creates)
replicates in order to engage in cognition. Though structurally dependent, the nervous system affords an innate plasticity and with appropriate alterations, it is conducive to learning and can adapt itself towards broader
interactions and human self-consciousness (Maturana and Varela, 1980a, 1980b).
Such broader interpretations of this theory
are addressed in Mingers (1991), such as the family nexus which via its idiosyncratic
behavioral and linguistic interactions (in relationship to hereditary factors,
socioeconomic status, environment, culture, etc.), creates a peculiar structured
reality that for the best part is only accessible to the members themselves, and through which a prospective psychotherapist
must explore in order to fathom out those recurrent patterns of conversation and behavior influencing the cognitive malfunctioning
of any concerned (cf Laing and Esterson, 1964).

There are related theories with varying degrees of overlap to the overall concepts of Maturana and Varela (1980a, 1980b).
One example is a biogenetic-structuralist theory (Laughlin and d'Aquili, 1974)
 where connections within an evolutionary context are made between neural organization/brain--functional activity and the environment.
 The result is ``acquired models of reality'', again, a cognitive outcome of the response to and engagement with the environment, proposing
a concept of `neurognosis', a kind of `holographic' model based upon biogenetically induced rudimentary information embedded in various associated regions of the brain, as taken from birth towards the later stages of maturity.
The neurognostic model depends upon ontogenetic feedback from an external reality according to which
acquired components are continually engaged with sensory input. Such modes are claimed to be
represented in neural processes by probability expectations which we shall see,
lie at the heart of information theory, and which encode the behavior of these process in relationship to the environment.
Thus it may be reasonable to suppose that enriched environments of some sort reciprocate to a greater advantage than others, and the
structures governing the alignment of the self to the environment to an extent evolved in accordance to some degree of
mastery over the latter. Take for instance, the hypothalamic--pituitary--adrenal (HPA) axis
(as governing the neurophysiology of the ``flight or fight''
mechanism) is cognitive in the sense of Atlan and Cohen (1998). If there is an arousal of the individual's close
environment, then mind, memory and emotional cognition engage,
evaluate and select appropriate responses. The HPA accelerates this
process and possible malfunctioning may induce hyper--reactivity
as observed in cases of post--traumatic stress disorder.
Depression, as another example,  may be partly viewed as the evolution of a structure conducive to a negative alignment
of the self with an external reality. If we think of a child as gradually
mastering its environment via a symbiotic relation with its mother, then take away the mother, the
inability to further manage the environment activates a negative kind of (neurognostic) structure such as those that have
been researched in the context of various evolutionary theories of depression that are founded upon the occurrence of
attachment-defeat-loss, diminished opportunities,
down-regulation of foraging capability, social/professional rank, etc. incurred with varying risk factors
within a culturally influenced environment (see e.g. Gilbert, 2006; Moffitt et al., 2006; Wallace, 2009).
Another example is to consider the body's
blood pressure control system consisting as a network of
cognitive systems which compare a set of incoming signals with
an internal reference structure in order to select a suitable level of
blood pressure from possible levels; hence as claimed in (Wallace and Fullilove, 2008; Wallace and Wallace, 2009)
 an elaborate tumor control strategy must be at least as cognitive as the immune system itself.

\subsection{Trail systems and Roman roads}

Thus granted that most (if not all) given classes of cognitive modules interact within their cultural
environment, we may proceed to consider what happens when the environment \emph{is}
the communication medium itself. For instance, in typical AI
laboratory models where multi-agent, inter-sensing systems function in local coordinated tasks
(often with no explicit communication between agents), the eventual net effect may induce a `shared memory'
(Cao et al., 1997; Krieger et al., 2000) where, for instance, it was claimed that more energetic
and efficient foraging tasks were typical of multi-agent (robotic) systems compared to individual agents, and the former
tended to produce behavioral patterns similar to those of `ant--like' decentralized control systems.
On the other hand, biological regulatory networks besides being susceptible to alterations in the environment
and/or intracellular conditions, may operate stochastically in varying degrees as seen along the pathways of
neuronal signalling transduction (Manninen et al., 2006) that provides an analogy motivating
part of this paper (see \S\ref{ito-process}).
A similar scenario is that of neuronal `Trail Systems' (TS) in Glade et al. (2009):
single wire and logical gates in a self--organized bioprocessor along which self--propelled
particles communicate via traces etched out in the environment, thus creating the TS. The claim of Glade et al. (2009)
is that such systems, although not precisely defined, are capable of programming and function on the
basis of a Turing machine. It suggests an evolutionary factor by which various living beings, by activating
their respective nervous systems, have trained themselves to use models like the TS towards evolution within
and possible mastery over their environment, by means of simulating `trail' signals. More from an information-theory viewpoint,
Wallace (2009b) envisages similar ideas to trail systems as reminiscent of `Roman roads': decision making and tasking
within small local communities, eventually creating `roads', manifestly inter--connected cognitive modules
having different time constraints, but eventually creating extensions of `local consciousness'.
A rate distortion argument applies here to account for the mutual crosstalk between different modules
using the homology of the rate distortion function with free energy as will be described later.
Next we proceed with some specifics.

%%%%%%%%%%%%%%%%%%%%%%%%%%%%%%%%%%%%%%%%%%%%%%%%%%%%%%%%%%%%%%%%%%%%%%%%%%%%%%%%%%%%%%%%%%%%%%%%%%%%%%%%

\section{Rate distortion and source entropy}\label{rd-theory}

\subsection{The rate distortion function}\label{rd-function}

As is well--known, distortion arises when there is a fast relay of information through
some channel which exceeds the latter's capacity. One of the principles of the Shannon theory
is that in order to reproduce a message transmitted from a source to a
receiver, it is necessary to know what sort of information should be
transmitted and how.
For the purpose of engineering a communication system, one needs to
figure out a suitable encoding/decoding system
once the nature of the channel is specified. Following Berger (1971),
we briefly recall some of the basic principles involved.

\textbf{Source encoder}: We may consider some output $x(t)$ emanating
from the source as projected to a finite set of preselected images,
namely, the space of possible source outputs is partitioned into a
set of \emph{equivalence classes} and the source encoder informs the
channel encoder of that class containing the particular source output
observed.
Once the channel encoder is informed that the source output belongs
to say, the $m$-th equivalence class, it transforms the corresponding waveform $\tilde{x}_m(t)$
across the channel.

\textbf{Source decoder}: Within the system is a cascade of a channel encoder and a source decoder.
The channel decoder receives a waveform $\tilde{y}(t)$ of a corresponding function $y(t)$ over some time interval and decides upon the nature of the message as transmitted. Then it sends its approximation $m'$ of the message number to the source decoder which in turn creates $y_{m'}(t)$ to register the system's estimate of $x(t)$ over that time interval.
Initially, we may think of $x(t)$ and $y(t)$ as `waveforms', but in our case, we consider these as
consisting  of a language with its own intrinsic grammar/syntax, as well as `meaning' -- to be made more specific
in \S\ref{meaningful}. Analogous considerations apply to the channel signals $\tilde{x}(t)$ and $\tilde{y}(t)$ (see Figure 2).

\begin{figure}
\begin{center}
\includegraphics[width=75mm]{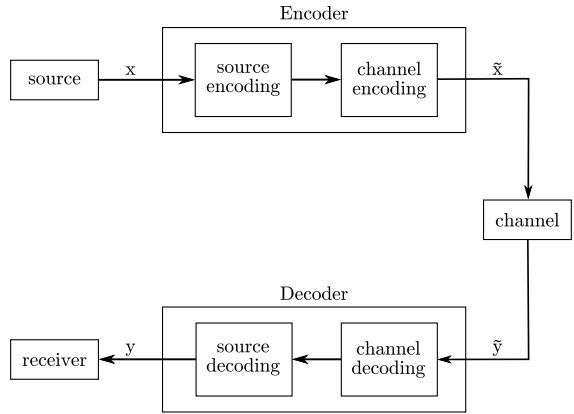}

\caption{Source-receiver encoding-decoding (as adapted from Berger, 1971, Fig. 1.2.1, p.4).}
\end{center}
\end{figure}

One of Shannon's notable results was that a communication system can be designed such that it achieves a
level of fidelity $D$ once the \emph{rate distortion} $R(D) \leq C$, where
$C$ denotes the channel capacity. Putting it another way, if the receiver can tolerate an average amount of distortion $D$, the rate distortion $R(D)$
is the effective rate at which the source can relay information with that level of tolerance.

The rate at which a source produces information subject to insisting upon perfect reproduction, is the
\emph{source entropy} $H$. Given a distortion measure such that perfect reproduction is assigned
zero distortion, then we have $R(0) = H$. As $D$ increases, $R(D)$ becomes a monotonically decreasing (convex) function which eventually is zero, typically at a maximum value for $D$ (see Berger, 1971, Chapter 1). This is a very basic observation, and typically in rate distortion theory
one seeks a reduction of $H$ by either slowing down the emission of coding, or encoding the relevant languages at a lower rate. In view of Shannon's theorem, as long as $C > H$, we will obtain appropriate fidelity in transmission. Inherent difficulties are clear, since the source rate may be corrupted due to low memory and coding congestion, hence the need for a communicating system to evolve so as recover the source data at the channel output satisfying the Shannon estimate.

\subsection{Average mutual information}\label{average}

Having mentioned the rate distortion function $R(D)$ we now follow Berger (1971) to give its specific definition in terms of \emph{average mutual information} (an alternative, and equivalent definition of $R(D)$ and a statement of the Rate Distortion Theorem will be given in Appendix \ref{appendix-B}). Firstly, for $k,j$ running over a suitable alphabet, let us write a given conditional probability assignment as $Q(k \vert j)$ such that in the usual way we have an associated joint distribution $P(j, k) = P(j) Q(k \vert j)$. We express \emph{the average distortion} as
\begin{equation}\label{av-1}
d(Q) = \sum_{j,k} P(j) Q(k \vert j) ~d(j,k),
\end{equation}
where $d(~,~)$ denotes the distortion measure. A conditional probability assignment $Q(k\vert j)$ is said to be \emph{$D$--admissible} if and only if $d(Q) \leq D$. The set of all $D$--admissible conditional probability assignments we denote by
\begin{equation}\label{av-2}
Q_D =\{ Q(k \vert j) : d(Q) \leq D \}.
\end{equation}
Along with an average distortion $d(Q)$, we also have an \emph{average mutual information}
\begin{equation}\label{av-3}
I(Q) = \sum_{j,k} P(j) Q(k \vert j) \log \big[\frac{Q(k\vert j)}{Q(k)} \big].
\end{equation}
Then for fixed $D$, the rate distortion function is defined as
\begin{equation}\label{av-4}
R(D) = \min_{Q \in Q_D} I(Q).
\end{equation}
Observe that if a parameter $s$ represents the slope of the function $R(D)$ at a point $(D_s,R_s)$
generated parametrically, we have $R'(D) = s$ (this is not exactly trivial: see Berger, 1971, Theorem 2.5.1).

\subsection{Meaningful paths}\label{meaningful}

More formally, a pattern of sensory input is mixed in an
unspecified but systematic algorithmic manner with a pattern of
internal ongoing activity to create a path of combined signals
$x=(a_{0}, a_{1},\ldots,a_{n}, \ldots)$.  Each $a_{k}$ thus
represents some functional composition of internal and external
signals. Wallace (2005) provides some neural network examples.

This path is fed into a highly nonlinear, but otherwise similarly
unspecified, decision oscillator, $h$, which generates an output
$h(x)$ that is an element of one of two disjoint sets $B_{0}$ and
$B_{1}$ of possible system responses. Let
\begin{equation}\label{response}
\begin{aligned}
B_{0} &\equiv b_{0}, \ldots, b_{k}, \\ B_{1} &\equiv b_{k+1},
\ldots,b_{m}.
\end{aligned}
\end{equation}

Assume a graded response, supposing that if
\begin{equation}
h(x) \in B_{0},
\end{equation}
the pattern is not recognized, and if
\begin{equation}
h(x) \in B_{1},
\end{equation}
the pattern is recognized, and some action $b_{j}, k+1 \leq j \leq
m$ takes place.
Such oscillators may be influenced by `forcing'
when a signal is subjected to some impulse such that its
frequency, and hence the response, adjusts accordingly with
respect to that applied impulse. More familiar oscillating
physical models react to this by exhibiting `beats' and
`resonance' for instance.

The principal objects of formal interest are paths $x$ which,
through information flow, trigger pattern
recognition-and-response. That is, given a fixed initial state
$a_{0}$, we examine all possible subsequent paths $x$ beginning
with $a_{0}$ and leading to the event $h(x) \in B_{1}$. Thus
$h(a_{0},\ldots,a_{j})\in B_{0}$ for all $0 < j < m$, but
$h(a_{0},\ldots,a_{m}) \in B_{1}$.

For each positive integer $n$,
let $N(n)$ be the number of high probability
grammatical/syntactical paths of length $n$ which begin with some
particular $a_{0}$ and further leading to the condition $h(x) \in
B_{1}$.  These are paths of combined signals as above, that are
structured to some language.
For short, we call such paths `meaningful', assuming, not
unreasonably, that $N(n)$ will be considerably less than the
number of all possible paths of length $n$ leading from $a_{0}$ to
the condition $h(x) \in B_{1}$.

One critical
assumption which permits an inference on the necessary conditions
constrained by the asymptotic limit theorems of information
theory, is that the finite limit
\begin{equation}\label{info1}
H \equiv \lim_{n \lra \infty} \frac{\log[N(n)]}{n},
\end{equation}
(the `uncertainty') both exists and is independent of the path
$x$. The rate distortion principle applies as follows (Wallace, 2005):
\emph{the restriction to meaningful
sequences of symbols increases the rate at which information can
be transmitted with arbitrary small error, and that the
grammar/syntax of the path can be associated with a dual
information source}. Here we may assume a typical information
source $\mathbf{X}$ to be `adiabatic', `piece-wise stationary' and
`ergodic' (APSE), and that a system engaging in a cognitive
process is describable as such. We list here the explanations:
\begin{itemize}
\item[(1)]
`Adiabatic' means that the changes are slow enough to allow the
necessary limit theorems to function.

\med
\item[(2)]
`Stationary' means that between pieces the probabilities hardly
change, and `piecewise' means that these properties hold between
phase transitions which are described using renormalization
methods (see Wallace, 2005).

\med
\item[(3)]
`Ergodic' means that in the long term, correlated sequences of
symbols are generated at an average rate equal to their (joint)
probabilities.
\end{itemize}
More specifically, the essence of `adiabatic' is that, when the
information source is parametrized according to some appropriate
scheme, within continuous `pieces' of that parametrization,
alterations in parameter values occur slowly enough so that the
information source $\mathbf{X}$ remains as close to stationary and
ergodic as needed to put to work the fundamental limit theorems of
information theory. In view of \eqref{info1}, the
Shannon uncertainty of $\mathbf{X}$ can be stated more specifically by (see e.g. Cover and Thomas, 1991):
\begin{equation}\label{info1a}
H[\mathbf{X}]= \lim_{n \lra \infty} \frac{\log[N(n)]}{n}.
\end{equation}.

\subsection{The fundamental homology}\label{feynman}

We recall how the information source uncertainty was defined as in
equation \eqref{info1}. This is quite analogous to the free energy
density of a physical system, equation \eqref{info5}, and the
relevance to a cognitive process can be explained by the following
steps. For instance, Feynman (1996) provides a series of
physical examples (based in part on the research of C. H. Bennett into the thermodynamics of
computing (Bennett, 1982) where this homology
is, in fact, an identity, at least for very simple systems.
Bennett (1982) argues, in terms of idealized irreducibly elementary
computing machines, that the information contained in a message
can be viewed as the work saved by not needing to recompute what
has been transmitted, or as Feynman (1996) puts it:
the information contained in a message is proportional to the amount of
free energy density needed to erase it.
The essential argument is that computing, in any form,
takes work. Thus the more complicated a cognitive process,
measured by its information source uncertainty, the greater its
energy consumption, and our ability to provide energy to the brain
is limited: typically, a unit of brain tissue consumes an order of
magnitude more energy than a unit of any other tissue.

The less information available to us concerning an event,
the higher is its entropy, and information retrieved is not without a cost in expenditure of energy, where
`cost' may be interpreted as the necessary number of bits needed to encode a message.
The thermodynamic minimum of energy in terms of bits of information is $k_B T
\log_2 e $ erg/bit ($= k_B T$ erg/nat). So
efficiency in an information system is essentially when there is the minimum
amount of energy expended in retrieving information. Specifically, if $F$ is taken to denote the free energy,
then taking $\Lambda$ is to denote the minimum number of nats/sec, the efficiency of the system
is given by $\eta = k_B TF^{-1} \Lambda$ (see e.g. Berger, 1971).

In a similar spirit to Bennett's work, Li and Vit\'{a}nyi (1992)
 consider the thermodynamic
costs of computation and how certain thermodynamic considerations can
give a recursively invariant notion of `cognitive distance' using a
kind of billiard dynamics approach.
In this case a minimal cognitive distance
between two objects will correspond to the minimal amount of work
expended for a given cognitive transformation of objects, either by
some computational procedure or by some neurocognitive function of the brain.
A higher descriptive level leading to more
complex and protracted algorithms, then leads to greater Kolmogorov complexity (Li and Vit\'{a}nyi, 1993).
As a particular application of the Bennett/Feynman ideas in the Global Workspace setting,
Wallace (2007)
argues that the cognitive disorder of `inattentional
blindness' emerges as a thermodynamic limit on processing capacity
in a topologically-fixed global workspace, i.e. one which has been
strongly configured about a particular task. Institutional and
machine generalizations seem clear.

%%%%%%%%%%%%%%%%%%%%%%%%%%%%%%%%%%%%%%%%%%%%%%%%%%%%%%%%%%%%%%%%%%%%%%%%%%%%

\section{Dynamic groupoids and their atlases}

\subsection{Concept of a groupoid}

Many cognitive processes exhibit the patterns of dynamical
systems (see e.g. Glazebrook and Wallace, 2009a). In such systems one aims to unify
the internal and external symmetries, and to be able to reduce
vast myriad--like network configurations into manageable schemes
involving the corresponding equivalence classes analogous to those
already mentioned in source encoding/decoding, etc. in \S\ref{rd-function} (see also
\S\ref{equiv-info} below). A precise way of doing
this lies within the categorical concept known as a \textit{groupoid}
(see Brown, 2006; Connes, 1994; Weinstein, 1996). In essence a groupoid $\grp$
consists of both a set of objects $X$ and a set of morphisms, or
`arrows', each of which project to an object in $X$, and all such
morphisms admit an inverse.

\begin{remark}The most familiar example of a
groupoid, as known to students of algebra, is that of a `group'
where there is a single object (`the identity').
Hence groupoids
can be viewed as extensions of the `group' concept to sets of
\textit{multiple identities} thus providing a wide scope of applications to
the dynamics of neurocognitive and socio--bioinformatic systems
(see e.g. Baianu et al., 2006; Glazebrook and Wallace, 2009a, 2009b; Golubitsky and Stewart, 2006;
Stewart et al., 2003; Wallace 2005; Wallace and Fullilove, 2008).
\end{remark}
A groupoid can be
depicted by
\begin{equation}
\a,\be~:~ \xymatrix{ \grp \ar@<1ex>[r]^{\a} \ar[r]_{\be} & X}
\end{equation}
where the groupoid morphisms $(\a, \be)$ onto objects, are called
the \emph{range} and \emph{source maps}, respectively. Informally,
the groupoid represents a feature of built in reciprocity between
its algebraic structures, internalizing and externalizing the
prevailing symmetries. The morphisms $\a, \be$ satisfy certain
algebraic relations of associativity, existence of two-sided
identities, etc. (details can be seen in e.g. Brown, 2006; Connes, 1994; Weinstein, 1996).
A groupoid can here be understood in relationship to a linkage by a
meaningful path of an information source dual to a cognitive
process for which the underlying principle is that: \emph{states
$a_{j}, a_{k}$ in a set $A$ are related by the groupoid morphism
if and only if there exists a high probability grammatical path
connecting them to the same base point, and the tuning across the
various possible ways in which that can happen -- the different
cognitive languages -- parametrizes the set of equivalence
relations and creates the groupoid.}

\begin{example}\label{ex2}
Since we have already mentioned equivalence classes in the context of source
encoding/decoding, it seems appropriate to see how an
equivalence relation $\R$ defined on (a set) $X$ takes shape as a groupoid.
Here we have the two projections $\alpha, \beta:\R \lra X$, and a product
$(x,y)(y,z)=(x,z)$ whenever $(x,y),(y,z) \in \R$ together with an
identity, namely $(x,x)$, for each $x \in X$.
Moreover, the essential equivalence relations (classes) derived
from a systems space (network) arise from the orbit equivalence
relation of some groupoid $\grp$ acting on that space (see e.g. Weinstein, 1996).
In the context of connected (sub)networks/graphs
which can reduced to equivalence classes, natural groupoid structures
come about in accordance with equivalence classes of relations $\R(xy)$,
as above, that is simply interpreted as having an edge linking node $x$ to node
$y$. Conversely, a groupoid (of equivalence relations) admits an underlying
graph structure via its implicit scheme of objects and morphisms
between objects (for details, see e.g. Brown, 2006).
Thus we have the two-way associations
whereby `objects' can be identified with `nodes', and `morphisms'
identified with `edges' in groupoids (of equivalence relations) and networks, respectively:$$
\begin{aligned}
\text{Network}~ &\ovsetLR{\text{equivalence relation}}
~\text{Groupoid} \\ \text{Network}~ &\ovsetLL{\text{underlying
graph}} ~\text{Groupoid}
\end{aligned}
$$
\end{example}

\subsection{Groupoid atlases}

An important observation about multi-tasking in
institutional and distributed cognitive systems concerns how the various
submodules interact. When a given subnetwork is represented in its groupoid form, then
such interactions naturally can be realized in terms of \emph{groupoid actions},
a topic that warrants further attention. Then we would like to see the overall
cognitive system in terms of such interacting groupoids and designed by
 an `atlas' of the latter, and one, as shown in Glazebrook and Wallace (2009a),
  containing the representation of several possible emergent
`giant components' induced by the outcome of local group(oid) actions within the
Workspace (see Figure 3). A workable concept seems to be that of a \emph{groupoid
atlas} (Bak et al., 2006) which provides a schematic representation
for coupling interactions between multi-agent systems and uses a
pasting together of the local dynamic groupoid actions with the
net effect of a `global' groupoid.

One commences from a family of dynamically interacting groupoids
$(\grp_{\As}) =\{ \grp_1, \grp_2, \ldots \}$ where each groupoid
has the same set of objects; this family is called a \emph{single
domain} or \emph{multiple groupoid}. \emph{A groupoid atlas} is
then defined as a set with a covering by patches each of which
comprise a single domain with global action, representing the
local processing  which is then globalized across the atlas.
This is a desirable effect
and one particularly suited to logically inscribing processors or
sensors (the `agents') within the cognitive modules of the
Workspace. As a descriptive mechanism, this atlas has the advantage of
admiting a weaker structure compared with that of a conventional
manifold since no condition of compatibility between
arbitrary overlaps of the patches is necessary. This is a key
property relevant to the structure of cognitive modules that can be
geared to equivalence class representations where flexibility in
the structure is a natural characteristic. In this way, the atlas provides
a convenient description of a web
of complexity representing the dynamic reciprocity of
tightly-knitted functional systems as was applied to small world
networks (Glazebrook and Wallace, 2009a). Time and space does not permit
including a mathematical outline of the construction; for the technical details we refer the reader to
Bak et al. (2006) and del Hoyo and Minian (2009). However, we intend to apply this concept with
some essential details in \S\ref{info-atlas} below.

\begin{figure}
\begin{center}
\includegraphics[width=30mm]{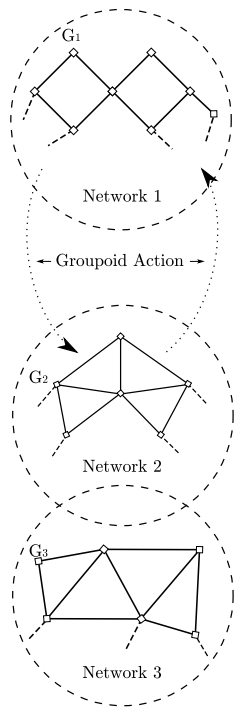}

\caption{Displayed are groupoids $\grp_1,\grp_2,\grp_3$ derived from the equivalence relations
of their respective networks of multi-agent type cognitive modules. They are viewed as the components of a groupoid atlas.
The groupoid actions are indicated by dotted arrows, in this case, between
$\grp_1$ and $\grp_2$, and
may represent the formation of network linkages as the system is shifted
via crosstalk, and for instance, how a `giant component' emerges
(see e.g. Glazebrook and Wallace, 2009a). This is a descriptive mechanism that could be seen, for instance, as enveloping
the underlying networks of the `reactive' system in Figure 1, and
is applicable to the content of \S\ref{info-atlas}. }
\end{center}
\end{figure}

%%%%%%%%%%%%%%%%%%%%%%%%%%%%%%%%%%%%%%%%%%%%%%%%%%%%%%%%%%%%%%%%%%

\section{Information and
the Onsager relations}\label{overview}

\subsection{A fundamental homology and the Onsager
relations}\label{fundamental1}

The information supported interactive cognitive
modules we have in mind are assumed to possess their own
internal metabolisms and mechanisms of
self-organization as reflective of vital biochemical processes. Just
as for the latter, the evolution of these `cognitive cells' is
characterized by reactive states of thermal nonequilibrium, in
accord with the laws of thermodynamics, and the capability to
assimilate information which we study in respect of source
uncertainty. This we can achieve by applying the Onsager relations
of nonequilibrium thermodynamics (Kurzynski, 2006; Landau and Lifshitz, 2007).
The reasoning starts by
observing how a fundamental homology between the information
source uncertainty dual to a cognitive process and the free energy
density of a physical system can arise mainly from the formal
similarity between their definitions in the asymptotic limit.

\subsection{The Groupoid Free Energy Density}

Recall that for a thermodynamic state of a given system at fixed temperature $T$ with energy $E$
and entropy $S$, the \emph{free energy density} $F$  is defined to be
\begin{equation}
F = E - TS.
\end{equation}
In the Hamiltonian formulism one takes the volume $V$ and the partition function
$Z(K)$ derived from the system's Hamiltonian at inverse temperature $K$
(Kurzynski, 2006; Landau and Lifshitz, 2007). The free energy density is then defined to be
\begin{equation}\label{info5}
\begin{aligned}
F[K] &= \lim_{V \lra \infty} -\frac{1}{K} \frac{\log[Z(K, V)]}{V} \\
&= \lim_{V \lra \infty} \frac{\log[\what{Z}(K, V)]}{V}, ~ \text{where}~ \what{Z}= Z^{-\frac{1}{K}}.
\end{aligned}
\end{equation}
Consider now an information source $H_{\grp_{\alpha}}$ over a corresponding groupoid
$\grp_{\alpha}$. The probability
of $H_{\grp_{\alpha}}$ is given by:
\begin{equation}
P(H_{\grp_{\alpha}}) = \frac{\exp[-H_{\grp_{\alpha}}K]}{\sum_{\beta} \exp[-H_{\grp_{\beta}}K]}~,
\end{equation}
where the normalizing sum is over all possible subgroupoids of the largest
available symmetry groupoid. Now let
\begin{equation}
Z_{\grp} = \sum_{\alpha} \exp [-H_{\grp_{\alpha}}].
\end{equation}
\emph{The groupoid free energy density
(GFE) of the system $F_{\grp}$ at inverse normalized
equivalent temperature $K$} is defined as
\begin{equation}
F_{\grp}[K] = - \frac{1}{K} \log [Z_{\grp}(K)].
\end{equation}
With each such groupoid $\grp_{\alpha}$ of the (large) cognitive
groupoid, we can associate
a dual information source $H_{\grp_{\alpha}}$. We recall the rate distortion function between the message sent by the cognitive process and the observed impact, while noting that both $H_{\grp_{\alpha}}$ and $R(D)$ may be considered as free energy density measures. In a sense, $R(D)$ constitutes a sort of `thermal bath'
for the process
of cognition. Then the probability of the dual information source can be expressed by
\begin{equation}
P(H_{\grp_{\alpha}}) =
 \frac{\exp[-H_{\grp_{\alpha}}/\kappa R(D) \tau]}{\sum_{\beta}
   \exp[-H_{\grp_{\beta}}/
 \kappa R(D) \tau]}~,
\end{equation}
where $\kappa$ denotes a suitable dimensionless constant characteristic of the system
in the context of a fixed machine response time $\tau$. The
sum is over all possible subgroupoids of the largest available symmetry groupoid. Accordingly,
the term $R(D) \kappa$ represents a `rate distortion energy', in this case,
a kind of `temperature analog'. In the context of a fixed $\tau$, a decline in $R(D)$
(on increase in average distortion), acts to `lower the machine temperature', driving it to more simple
and less less rich behaviors.

\subsection{A groupoid atlas of information sources}\label{info-atlas}

The groupoids $\grp_{\alpha}$ can indeed be taken to comprise a groupoid atlas ${\As}$
on an appropriate set $X_\As$. In Bak et al. (2006) this is motivated by
considering a group $G$ in the standard algebraic sense, and a family
\begin{equation}
\begin{aligned}
~&  \{ (G_\As)_{\a} \curvearrowright (X_\As)_{\a} : \a \in
\Psi_\As \} \\ &= \{ (G_\As)_{\a} \times (X_\As)_{\a} \lra
(X_\As)_{\a} : \a \in \Psi_\As \},
\end{aligned}
\end{equation}
of group actions `$\curvearrowright$' on subsets $(X_\As)_{\a}
\subseteq X_\As$, where the local groups $(G_{\As})_{\a}$ and the
corresponding subsets $(X_\As)_{\a}$ are indexed by an indexing
set $\Psi_\As$ called \emph{the coordinate system of $\As$} which is seen to satisfy
certain conditions (Bak et al., 2006). Now the family of local groups can be replaced by
a family of local groupoids $(\grp_{\As})$ defined with respective object sets $(X_\As)_{\a}$, and with
a coordinate systems $\Psi_{\As}$ that is equipped with a reflexive relation denoted by
$\leq$. This data is to satisfy the following conditions (Bak et al., 2006):

\begin{itemize}
\item[(1)]
If $\a \leq \be$ in $\Psi_{\As}$, then $(X_\As)_{\a} \cap
(X_\As)_{\be}$ is a union of components of $(\grp_{\As})$, that
is, if $x \in (X_\As)_{\a} \cap (X_\As)_{\be}$ and $g \in
(\grp_\As)_{\a}$ acts as $g : x \lra y$, then $y \in (X_\As)_{\a}
\cap (X_\As)_{\be}$.

\item[(2)]
If $\a \leq \be$ in $\Psi_{\As}$, then there is a groupoid
morphism defined between the restrictions of the local groupoids
to intersections
\begin{equation}
\phi^{\be}_{\a}: (\grp_\As)_{\a} \vert (X_\As)_{\a} \cap
(X_\As)_{\be} \lra (\grp_\As)_{\be} \vert (X_\As)_{\a} \cap
(X_\As)_{\be},
\end{equation}
and which is the identity morphism on objects.
\end{itemize}
Thus each of the $\grp_{\alpha}$ with its associated dual information
source $H_{\grp_{\alpha}}$ constitutes a component of an atlas which incorporates
the dynamics of an (inter)reactive system through information sources, by means of the intrinsic (groupoid) actions.
For simplicity, let us refer to this atlas as $\As$. Suppose we have another such atlas $\Bs$ representing a
separate system which can be related to $\As$ via a suitable transformation. To make matters precise
we consider then a morphism $f: \As \lra \Bs$ prescribed by a triple $(X_f, \Phi_f, \grp_f)$ satisfying (del Hoyo and Minian, 2009):
\begin{itemize}
\item[(1)] $X_f: X_{\As} \lra X_{\Bs}$ is a set-theoretic function.

\item[(2)] $\Phi_f: \Phi_{\As} \lra \Phi_{\Bs}$ is a function that preserves the relation $\leq$.

\item[(3)] $\grp_f: \grp_{\As} \lra \grp_{\Bs}$ is a (generalized) natural transformation of groupoid diagrams over the function
$\Phi_f$ which restricts to $X_f$ on objects.
\end{itemize}
These conditions can summarized in the following straightforward way. For each $\a$, a function $\grp_{f(\a)}: \grp_{\a} \lra \grp_{\Phi_f(\a)}$ is
given, such that for objects $\obj(\grp_f(\a)) = X_f \vert X_{\a}$, and if $\a \leq \be$, the diagram
\begin{equation}
\begin{CD}
\grp_{\a} @>>> \grp_{\Phi_f(\a)}
\\ @VVV @VVV\\
\grp_{\be} @>>> \grp_{\Phi_f(\be)}
\end{CD}
\end{equation}
is commutative.

\subsection{Biochemical data compression}\label{multiphase}

Further motivation is provided by considering phases as chemically and thermodynamically homogeneous
when formally compared with average mutual information such as applied to living systems that are capable
of assimilating and using information reaped either from their environment or from information that is intrinsic
to their particular system via genetic characteristics. Here we reflect in part upon the theme of \S\ref{feynman} in which we discussed
certain physical means by which information runs at the cost of expending free energy. On course with the fundamental
homology, it befits us to consider some physical equations that are homologous to the variational calculus of the rate
distortion function, and follow to some extent Berger (1971, \S6.4). The idea is that when a living information system is confronted
with an environment, it reacts towards it at an atomic-molecular level.
Given then some thermodynamic system, let $n_{jk} $ be the number of atomic weights of substance $j$ that end up
in phase $k$, and let $n_j$ be the number of atomic weights that were introduced originally, so that
$n_j = \sum_k n_{jk}$. The multiphase (chemical) equilibrium problem
involves determining the $n_{jk}$. To proceed, let us express the free energy as $F= F_1 + F_2$, where
\begin{equation}\label{mp-1}
\begin{aligned}
F_1 &= \sum_{j,k} n_{jk} c_{jk}, \\
F_2 &= \sum_{j,k} n_{jk} \log \big[ \frac{ n_{jk}}{\sum_j n_{jk}} \big],
\end{aligned}
\end{equation}
where the $c_{jk}$ are \emph{the free energy constants},
and where the term contained in the logarithm is the chemical potential of the
reactant $j$ in phase $k$. Let now $n = n_0 + \cdots + n_{M-1}$ (for some $M$), ~ $P_j = n_j/n, ~
Q_{k \vert j} = n_{jk}/n_j$ and $Q_k = \sum_i n_{jk}/n$, then
\begin{equation}\label{mp-2}
F = F_1 + F_2 = n \sum_{j,k} P_j Q_{k \vert j} \big[ c_{ij} + \log \big[ \frac{Q_{k \vert j}}{Q_k} \big]
\big ] - n \sum_j P_j \log P_j.
\end{equation}
Since in principle the $n$ and $P_j$ can be determined, the minimization of $F$ reduces to minimizing the
double summation on the right side of \eqref{mp-2}. Letting $d_{jk} = - c_{jk}/s$, then `nature' selects
the $n_{jk}$ so as to minimize the quantity
\begin{equation}\label{mp-3}
\mathcal V = \sum_{j,k} P_j Q_{k \vert j} \big[ \log \big[ \frac{Q_{k \vert j}}{Q_k}\big] -s d_{jk} \big].
\end{equation}
The crucial observation here is that \eqref{mp-3} is formally the same as
\begin{equation}\label{mp-4}
\mathcal V = I(Q) - s d(Q),
\end{equation}
in other words, the function to be minimized is the difference between the average mutual information and $s$ times the average distortion. This is minimized by an appropriate choice of $Q=Q(k \vert j)$ such as to determine a point on the $R(D)$ curve where, as in \S\ref{average}, we have $R'(D) =s$. The multiphase chemical equilibrium is a
 local process; the overall system itself may in general remain in a state far-from-equilibrium.

According to Berger (1971), `nature' then automatically performs the analogous minimization
in multiphase chemical equilibrium, and when an information system encounters an environment, the reaction
on the molecular level follows these principles. Thus the free energy of the combined system and the
environment is minimized, although this is at the cost of the system's free energy capacity where there
will inevitably be some heat dissipation. Eventually the system can fine-tune its coding mechanism
and re-configures itself for the task of gaining
more specialized knowledge about its environment, and thanks to newly acquired information, it may engage the latter
 to its advantage along with detecting other drifting cells of information. How such optimizing procedures can be realized
 in an evolutionary context, is of course one of the main tasks of applying rate distortion arguments
 (cf Wallace and Wallace, 1998, 1999, 2008, 2009). For instance, a scenario studied in Tlusty (2007, 2008)
 concerns how the genetic code maps the 64 nucleotide triplets (codons) to
 20 amino acids. In terms of this mapping, the code is viewed as a noisy information channel and the claim is that
 evolutionary characteristics determine the emergence of the code via appropriate selection of amino acids which minimise
 the risk of errors, and subsequently the code emerges at a `supercritical' phase transition once the mapping ceases
 to be random.

%%%%%%%%%%%%%%%%%%%%%%%%%%%%%%%%%%%%%%%%%%%%%%%%%%%%%%%%%%%%%%%%%%%%%%%%%%%%%%%%%%%%

\section{The Onsager relations in the context of information}

\subsection{The basic equations}

Understanding the time dynamics of cognitive systems away from
phase transition critical points thus requires a phenomenology
similar to the Onsager relations. If the dual source uncertainty
of a cognitive process is parametrized by some vector of
quantities $\mathbf{K} \equiv (K_{1}, \ldots, K_{m})$, then, in
analogy with nonequilibrium thermodynamics, the gradients in the
$K_{j}$ of the \textit{disorder}, defined as
\begin{equation}\label{info6}
S \equiv H(\mathbf{K})-\sum_{j=1}^{m}K_{j}~ \partial H/\partial
K_{j},
\end{equation}
become of central interest. Equation \eqref{info6} is similar to
the definition of entropy in terms of the free energy density of a
physical system, as suggested by the homology between free energy
density and information source uncertainty described above.
Pursuing the homology further, the generalized Onsager relations
defining temporal dynamics become
\begin{equation}\label{info7}
 dK_{j}/dt = \sum_{i} L_{ji}~ \partial S/\partial K_{i},
\end{equation}
where the (kinetic coefficients) $L_{ji}$ are, in first order, constants
interpreted as reflecting the
nature of the underlying cognitive phenomena.
The partial derivatives $\partial S/\partial K$ are
analogous to thermodynamic forces in a chemical system, and may be
subject to override by external physiological driving mechanisms
as shown in (Wallace, 2005; Wallace and Fullilove, 2008) along with further extensions
of these dynamical procedures.

\begin{remark}
Equation \eqref{info7} is `general' in the sense that we do not
necessarily assume the symmetry condition
$L_{ji} = L_{ij}$ which in this latter case expresses
Onsager's 4-th law of thermodynamics (see e.g. (3.7) in
Kurzynski, 2006). The matrix $L = [L_{ij}]$ is to
be viewed empirically, in the same spirit as the slope and
intercept of a regression model, and may have a structure far
different when compared to the more basic, more familiar chemical or physical
processes. Generally, information sources are notoriously one--way in time,
as exemplified by the patent linguistic scarcity of palindromic structures
that do actually make some sense.
\end{remark}

\subsection{Equivalence classes of information sources}\label{equiv-info}

Equations \eqref{info6} and \eqref{info7} can be derived in a
simple parameter-free covariant manner which relies on the
underlying topology of the information source space implicit to
a process. Different cognitive phenomena have, according to our
development, dual information sources, and we are interested in
the local properties of the system near a particular reference
state. We impose a topology on the system, so that, near a
particular `language' $A$, dual to an underlying cognitive
process, there is (in some sense) an open set $U$ of closely
similar languages $\hat{A}$, such that $A$ and $\hat{A}$ are
subsets of $U$. Note that it may be necessary to coarse-grain the
system's responses to define these information sources. The
problem is to proceed in such a way as to preserve the underlying
essential topology, while eliminating `high frequency noise'. The
formal tools for this can be found in e.g. Burago et al. (2001).

Since the information sources dual to the cognitive processes are
similar, for all pairs of languages $A, \hat{A}$ in $U$, it is
possible to make use of the following:

\begin{itemize}
\item[(1)]
Create an embedding alphabet which
includes all symbols allowed to both of them.

\med
\item[(2)]
Define an information-theoretic distortion measure in that
extended, joint alphabet between any meaningful (i.e. high
probability grammatical/syntactical) paths in $A$ and $\hat{A}$,
which we write as $d(Ax, \hat{A}x)$ (Ash, 1990; Cover and Thomas, 1991). Note that these
languages do not interact, in this approximation.

\med
\item[(3)]
Define a metric on $U$, for example,
\begin{equation}\label{info8}
\mathcal{M}(A, \hat{A}) = \vert \lim \frac{\int_{A,\hat{A}} d(Ax,
\hat{A}x)}{\int_{A,A} d(Ax,A\hat{x})} - 1 \vert,
\end{equation}
using an appropriate integration limit argument over the high
probability paths. The usual metric properties apply as in Burago et al. (2001).
\end{itemize}

Note that these conditions can be used to define equivalence
classes of \textit{languages}, where previously we defined
equivalence classes of \textit{states} which could be linked by
meaningful paths to some base point. This led to the
characterization of different information sources from which a
formal topological manifold, \emph{which is an equivalence class
of information sources}, can be constructed. As a working hypothesis we may assume
this to be a standard differentiable manifold in which the set of such
equivalence classes generates the \textit{dynamical groupoid} (cf Glazebrook and Wallace, 2009a, 2009b),
and then study those
mechanisms, internal or external, which can break that groupoid
symmetry. In particular, the imposition of a metric structure on
this groupoid, and on its base set, would permit a nontrivial
interaction between orbit equivalence relations and isotropy
groups, leading to interesting algebraic structures.

Since $H$ and $\mathcal{M}$ are both scalars, a `covariant'
derivative can be defined directly as
\begin{equation}\label{info9}
dH/d\mathcal{M} = \lim_{\hat{A} \lra A}
\frac{H(A)-H(\hat{A})}{\mathcal{M}(A,\hat{A})},
\end{equation}
where $H(A)$ is the source uncertainty of language $A$. Suppose
the system is set in some reference configuration $A_{0}$. To
obtain the unperturbed dynamics of that state, impose a Legendre
transform using this derivative, defining another scalar
\begin{equation}\label{info10}
S \equiv H - \mathcal{M} dH/d\mathcal{M}.
\end{equation}
The simplest possible Onsager relation -- here seen as an
empirical, fitted, equation like a regression model -- in this
case becomes
\begin{equation}\label{info11}
d\mathcal{M}/dt = L dS/d\mathcal{M},
\end{equation}
where $t$ is the time and $dS/d\mathcal{M}$ represents an analog
to the thermodynamic force in a chemical system (cf \S\ref{multiphase}).
Relevant here are patterns of oscillatory-like behavior where a weak
signal is amplified by the presence of noise as result of some
synchronized hopping around local extrema. The standard
terminology for this phenomenon is \emph{stochastic resonance}
(Gammaitoni et al., 1998) and we will proceed to give some idea of how this can
be related to certain types of cognitive processes. Since we are
working with stochastic differential equations, the first step is
to modify the equation of thermodynamic force accordingly. To this
extent equation \eqref{info11} is rewritten as
\begin{equation}\label{info12}
d\mathcal{M}/dt = L dS/d\mathcal{M} + \sigma W(t),
\end{equation}
where $\sigma$ is a constant and $W(t)$ represents a white noise
term. Again, the quantity $S$ is seen as a function of the
parameter $\mathcal{M}$. This leads directly to a family of
classic stochastic differential equations expressed as
differential 1-forms
\begin{equation}\label{info13}
d\mathcal{M}_{t}= L(t, dS/d\mathcal{M})~dt + \sigma(t,
dS/d\mathcal{M})~dB_{t},
\end{equation}
where $L$ and $\sigma$ are appropriately regular functions of $t$
and $\mathcal{M}$, and $dB_{t}$ represents the noise structure.

Such cognitive-epigenetic systems which are driven by stochastic and noise driven diffusion processes, may be
suitably conditioned to admitting further noise perturbations that lead to a degree of stochastic resonance
capable of amplifying a relatively weak signal or actually reducing the level
of randomness in the system. Such resonance may function as a catalyst towards the system's
self-organization and complexity, in the same way as open systems far-from-equilibrium require
internal amplification in order to reach a macroscopic dynamical structure (Gammaitoni et al., 1998; Prigogine, 1980; West
et al., 2005).

\subsection{Rate distortion dynamics}

Recall that the rate distortion function $R(D)$ defines the minimum channel capacity necessary for the system to have an average
distortion $\leq D$, thus imposing a limit on the information source uncertainty and suggesting how
distortion measures can drive information system dynamics. In other words, $R(D)$ affords a homological
relation to free energy density, very much along the lines of the above relation between free energy
density and information source uncertainty. Accordingly, it is proposed that the dynamics of cognitive modules
interacting in characteristic real--time $\tau$ will be constrained by the system as described in terms of
$R(D)$, but now we generalize matters as in Wallace and Wallace (2008) by producing a vector--valued function $R(\mathbf{Q})$ where in the vector
$\mathbf{Q}= (Q_1, \ldots, Q_k)$ the first component is defined to be the average distortion, and then (cf \eqref{info6})
\begin{equation}\label{rdd-1}
S_R \equiv R(\mathbf{Q})-\sum_{i=1}^{m}Q_{i}~ \partial R/\partial
Q_{i},
\end{equation}
which leads to the deterministic and stochastic systems of equations analogous
to the Onsager relations of nonequilibrium thermodynamics
\begin{equation}\label{rdd-2}
 dQ_{j}/dt = \sum_{i} L_{ji}~ \partial S_R/\partial Q_{i},
\end{equation}
together with
\begin{equation}\label{rdd-3}
dQ^j_t = L^j(Q_1, \ldots, Q_k, t)~dt + \sum_i \sigma^{ji}(Q_1, \ldots, Q_k, t) ~ dB^i_t,
\end{equation}
where the $dB^i_t$ represents often highly structured stochastic noise whose properties may be described
in terms of Brownian motion and quadratic variation (see e.g. Kunita, 1990; Protter, 1995).

At this stage we introduce several examples for which part of the purpose will be to
motivate introduction of the It\^{o} principle which we will do so below.

\begin{example}\label{rdd-ex1}
Firstly, for a simple Gaussian channel with noise having zero mean and variance $\sigma^2$,
we have
\begin{equation}\label{rdd-4}
S_R = R(D) - D dR(D) /dD = \frac{1}{2} \log (\sigma^2/D) + \frac{1}{2}.
\end{equation}
The simplest possible Onsager relation becomes
\begin{equation}\label{rdd-5}
dD/dt = - \mu dS_R/dD = \frac{\mu}{2D},
\end{equation}
in which the term $- dS_R/dD$ represents the force of an `entropic wind' which is a kind of
internal dissipation inevitably driving the real--time system of
interacting cognitive information sources toward greater distortion. Equation \eqref{rdd-5}
has a solution $D = \sqrt{\mu t}$, showing in this case that the average distortion increases
monotonically with time. Following Wallace (2009a, \S7.2), this example shows
that such a system will inevitably succumb to a relentless entropic force, requiring
\emph{a constant free energy expenditure for maintenance of some fixed average distortion
within the system's communication between them}. The distortion in this case will, without free energy input,
have a time dependence $D = f(t)$, with $f(t)$ montonically increasing in $t$, eventually
leading to the punctuated failure of the system. Further, in the Einstein diffusion equation,
a straightforward argument of Wallace (2009a, \S7.2)shows that the standard deviation of the particle position
increases in proportion to $\mu t$. Thus, whereas we do not expect the high correlations of an information
source to exhibit typical Brownian motion, it does seem to be the case that the \emph{distortion} in communication
between the interacting cognitive modules within the appropriate context of the Onsager relations, does display
Brownian motion which may be of bounded variation in certain cases.
\end{example}

\subsection{Rate distortion coevolutionary dynamics}

Here we consider different cognitive developmental subprocesses of gene expression characterized by
information sources $H_m$ interacting through chemical or other signals, and assume \emph{that different processes become each other's principal environments} which is a suitable hypothesis within a broad coevolutionary context.
Let
\begin{equation}\label{co-ev1}
H_m = H_m (K_1, \ldots, K_s, \ldots, H_j, \ldots ),
\end{equation}
where the $K_s$ represent other relevant parameters, and $j \neq m$. We regard the dynamics of this system as driven by a recursive network of stochastic differential equations.
Letting the $K_j$ and $H_m$ all be represented as parameters $Q_j$ (with the caveat that $H_m$ does not depend on itself), we follow the generalized Onsager formulation of Wallace and Wallace (2009), in terms of the equation
\begin{equation}\label{co-ev2}
S^m = H_m - \sum_i Q_i ~\del H_m /\del Q_i,
\end{equation}
to obtain a recursive system of \emph{phenomenological Onsager relations}, in terms of a system of stochastic differential equations
\begin{equation}\label{co-ev3}
dQ^j_t = \sum_i ~[ L_{ji}(t, \ldots,  \del S^m / \del Q^i, \ldots)~dt + \sigma_{ji}(t,
\ldots, \del S^m/ \del Q^i, \ldots )~dB^i_{t}],
\end{equation}
in which, for ease of notation, both the terms $H_j$ and the external $K_j$'s are expressed
by the same symbol $Q_j$. As $m$ ranges over the $H_m$ we could allow different kinds of
`noise' $dB^i_t$, having particular forms of quadratic variation which may represent
a projection of environmental factors within the scope of a rate distortion manifold (Glazebrook and Wallace, 2009b).

The next step in extending \eqref{co-ev3} is to bring in rate distortion functions for mutual crosstalk between a set of interacting cognitive modules by using the homology of $R(D)$ itself. To this extent, consider different cognitive processes indexed $1, \ldots, s$, and take the mutual rate distortion functions $R_{ij}$ characterizing communication (and distortion) between them. At the same time the essential
parameters remain the characteristic time constants of each process, $\tau_j$, for $1 \leq j \leq s$,
together with an overall embedding free energy density $F$.
Taking the $Q^{\a}$ to run over all the relevant parameters and mutual rate distortion functions (along with the distortion measures $D_{ij}$),
then \eqref{co-ev2} now takes shape as
\begin{equation}
S^{ij}_R = R_{ij} - \sum_k Q^k ~ \del R_{ij} / \del Q_k,
\end{equation}
and accordingly \eqref{co-ev3} becomes
\begin{equation}\label{co-ev5}
dQ^{\a}_t = \sum_{\be = \{ij\}} [ L_{\be}(t, \ldots,  \del S^{\be}_R/ \del Q^{\a}, \ldots)~dt + \sigma_{\be}(t,
\ldots, \del S^{\be}_R / \del Q^{\a}, \ldots )~dB^{\be}_{t}].
\end{equation}
This last equation generalizes the treatment in terms of crosstalk, its distortion, the inherent time
constants of the different cognitive modules, and \emph{the overall available free energy density}.

\begin{example}\label{coev-ex1}
For a Gaussian channel and fixed embedded communication free energy density $F$ representing the
richness of incoming information from the interacting cognitive modules, we extend \eqref{rdd-5} to
\begin{equation}\label{co-ex-1}
dD/dt = \frac{\mu}{2D} -\a F, ~ \a >0,
\end{equation}
that represents the communication distortion between the modules. The equilibrium solution is
$D_{\text{equil}} = \frac{\mu}{2 \a F}$.
The difference between \eqref{rdd-5} and \eqref{co-ex-1} is that whereas in the former case, the distortion
grows directly as the square root of the elapsed time, equation \eqref{co-ex-1}
reveals there is a finite, equilibrium, average distortion that is inversely proportional to the available
environmental or informational free energy that the interacting systems can implement in order to
navigate their actions.

The above situation can be generalized to $D_{\text{equil}} = \frac{1}{g(F)}$, where $g(F)$ is monotonically
increasing in $F$. On introducing  a characteristic response time variable $\tau$, so that
\begin{equation}\label{co-ex-2}
dD/dt = \frac{\mu}{2D} - g(F) h(\tau),
\end{equation}
where $h(\tau)$ is also monotonically increasing, leads to
\begin{equation}\label{co-ex-3}
D_{\text{equil}} = \frac{\mu}{2g(F)h(\tau)}.
\end{equation}
This example reveals that given a fixed rate of available information free energy, the increasing
allowable response time \emph{decreases} average distortion in the interaction.
\end{example}

\begin{example}\label{coev-ex2}
Suppose now that feedback is allowed so that the system actively seeks information in proportion to
the distortion between intent and impact, then the Onsager relation for a Gaussian channel becomes
\begin{equation}\label{co-ex-4}
dD/dt = \frac{\mu}{2D} - g(F) h(\tau) D,
\end{equation}
and
\begin{equation}\label{co-ex-5}
D_{\text{equil}} = (\frac{\mu}{2g(F)h(\tau)})^{\frac{1}{2}},
\end{equation}
which is significantly smaller than \eqref{co-ex-3}, and is effectively the classic result for Brownian motion
in a harmonic central field (e.g. equation (54) of Wang and Uhlenbeck, 1945).
\end{example}

\subsection{Multidimensional It\^{o} process}\label{ito-process}

Together with the multiphase equilibrium problem of
\S\ref{multiphase} we have so far pursued a theme of how the stochastic simulation of
biochemical systems closely parallels that of evolutionary--genetic systems.
An initial observation here, and one that will motivate further ideas, is that a
stochastic differential equation of the type \eqref{co-ev5} should in principle model the dynamics of large,
intricate networks that are constrained by the costs of actual computational time, which seems relevant in certain senses
to many epigenetic processes.
The general setting for such processes often involves that of an It\^{o} stochastic DE (viz \emph{It\^{o} process}) and this is how we intend
to view \eqref{co-ev5} as a kind of `master equation'.

For the readers sake, we remark why such a level of mathematical formality is necessary
by recalling a basic difficulty: the sample paths $B^{\be}_t$ of Brownian motion are not in general functions of
bounded variation, so that $dB^{\be}_t$ is not defined as for that of the usual Riemann-Stieltjes integral.
One may start by supposing that for almost all samples, $Q_t$ in \eqref{co-ev5}
is independent of future Brownian motion $B_u^{\be} - B_t^{\be}, ~ u \geq t$;
otherwise said $Q^{\a}_t$ is an \emph{adapted} stochastic process.
Putting it another way, the information available at a given time includes the history of the process
at that time. More generally, the stochastic process may be taken to be \emph{predictable}
in order to define the It\^{o} integral of the equation (see e.g. \S2.3 of Kunita, 1990 and Appendix
\ref{appendix-A} here).

On the other hand, there is part justification for making assumptions of (local) boundedness of
variation in the Brownian motion incurred via distortion, and
in particular, the adapted condition, points to those examples and observations
already quoted that concern the dispensation of available free energy. Firstly,
in view of Example \ref{rdd-ex1} we may expect a constant free energy expenditure for maintenance of
some fixed average distortion in communication between the interacting cognitive modules. Secondly,
given a fixed rate of available information free energy, the increasing
allowable response time \emph{decreases} average distortion in the interaction (Example \ref{coev-ex1}),
and thirdly, the possibility that finite, equilibrium, average distortion is actually inversely
proportional to the available environmental or informational free energy that the interacting systems
can utilize (Example \ref{coev-ex2}). However, to make matters precise, it is appropriate to consider the formalities
of some filtered probability space $(\Omega, \bF, P)$, where $\bF = \{\F_t : t \geq 0 \}$
 (see Appendix \ref{appendix-A}) and postulate a multidimensional stochastic
 process given by the It\^{o} integral of the master equation \eqref{co-ev5}:
\begin{equation}\label{co-ev6}
Q^{\a}_t = Q^{\a}_0 + \sum_{\be = \{ij\}} [ \int^t_0  L_{\be}(s, \ldots,  \del S^{\be}_R/ \del Q^{\a}, \ldots)~ds + \int^t_0 \sigma_{\be}(s,
\ldots, \del S^{\be}_R / \del Q^{\a}, \ldots )~dB^{\be}_{s}].
\end{equation}
In order to state the conditions for which this process is well-defined, we first express \eqref{co-ev6}
in the simplified form
\begin{equation}\label{co-ev6a}
Q^{\a}_t = Q^{\a}_0 + A^{\a}(t) + \int^t_0 \sigma_{\be}~dB^{\be}_{s}.
\end{equation}
Then following Harrison (1985, Chapter 4):
\begin{itemize}
\item[(1)] $Q^{\a}_0$ is measurable with respect to $F_0$.

\item[(2)] $\sigma_{\be}$ is an adapted stochastic process, and $P\{\int^t_0 \sigma^2_{\be}(s)~ds < \infty \} = 1$, for all $t\geq 0$.

\item[(3)] The integral of `drift', $A^{\a}(t) = \int^t_0  L_{\be}~ds$, is a continuous and adapted variation-finite (VF) process.
\end{itemize}
Granted that our observations about the local boundedness of the (rate distortion) Brownian motion as essentially fulfilling these conditions, we now have an explicit stochastic model for the role of cross-talk, its distortion, the inherent time constraints
of the different cognitive modules, as well as the overall available free energy density,
where the $Q$ parameter structure represents the full--scale fragmentation of the system in the presence of some Weiner noise.
\begin{remark}
A further possible generalization of \eqref{co-ev5} is to introduce into that expression
a matrix valued function
$V_{\be} : \bR^m_{+} \lra \bR^{n \times n}$ to describe the intrinsic reaction rates (cf Manninen et al., 2006):
$$
dQ^{\a}_t = \sum_{\be = \{ij\}} [ L_{\be}(t, \ldots,  \del S^{\be}_R/ \del Q^{\a}, \ldots)~dt + V_{\be} \sigma_{\be}(t,x
\ldots, \del S^{\be}_R / \del Q^{\a}, \ldots )~dB^{\be}_{t}].
$$
\end{remark}

\subsection{The stochastic flow}

Towards the possibility of a stochastic flow generated by \eqref{co-ev5} (such as a Brownian flow of diffeomorphisms), we opt to simplify \eqref{co-ev5} accordingly. Following Kunita (1990) we write
\eqref{co-ev5} in a simplified form as
\begin{equation}\label{co-ev7}
dQ_t = \sum_{\a} dQ^{\a}_t = f_0 (Q_t, \del S_R/\del Q, t) ~dt + \sum^m_{k=1} f_k (Q_t, \del S_R/\del Q, t) ~dB^k_t.
\end{equation}
Typically, we would seek a solution starting from some $x$ at some time $s$; let us call these solutions
$Q_{s,t}$. Then a stochastic flow can be represented as the solution of a stochastic differential equation
of the type
\begin{equation}\label{co-ev8}
Q_{s,t}(x) = x + \int^t_s F(Q_{s,r}(x), \del S_R/\del Q, dr),
\end{equation}
where in this case, Brownian motion $F(x, \del S_R / \del Q, t) $ valued in vector fields, is given by
\begin{equation}\label{co-ev9}
F(x, \del S_R / \del Q, t) = \int^t_0 f_0(x, \del S_R/\del Q, r)~dr +
\sum^m_{k=1} \int^t_0 f_k (x, \del S_R/\del Q, r) ~dB^k_r.
\end{equation}
The formal conditions for \eqref{co-ev8} to produce a Brownian flow of diffeomorphisms are (Kunita, 1990):
\begin{itemize}
\item[(1)] $Q_{s,t}(x)$ is continuous in $s,t,x$.

\item[(2)] The map $Q_{s,t}(x): \bR^d \lra \bR^d$ (for suitable $d$) is a diffeomorphism for any
$s < t$.

\item[(3)] $Q_{s,u} = Q_{t,u}(Q_{s,t})$, for any $s < t < u$.
\end{itemize}
Note that conditions (1)-(3) are `almost everywhere (surely)' conditions. Also, such a flow generates a
\emph{holonomy} or \emph{geometric phase (transition)} which can be explained by the process of tracking internal states
in relationship to a spatiotemporal orientation. In more precise differential-geometric terms, holonomy results from
the parallel transport of vectors around a closed path, thus leading to a representation of the space
of the latter into a group of global symmetries.
The procedure for constructing a
holonomy groupoid associated to this flow concerns some mathematical technicalities, but is nevertheless standard
(see e.g. Connes, 1994; Moerdijk and Mr\v{c}un, 2003). Whereas this case is fairly well tempered, we would in general expect `singularities'
in the flow. The holonomy groupoid can be still be constructed, but this involves deeper mathematics outside of the scope of this paper
 (see e.g. Debord (2001) for details).

%%%%%%%%%%%%%%%%%%%%%%%%%%%%%%%%%%%%%%%%%%%%%%%%%%%%%%%%%%%%%%%%%%%%%%%

\section{Discussion and conclusions}

We have given here a descriptive account of cognitive modules as components of
epigenetic-evolutionary systems (as far-from-equilibrium open systems)
embedded within the context of environment and culture. Using dynamic groupoids of network
equivalence classes we have put into atlas form the various constituent (inter)reactive
systems based on rate distortion principles of the Shannon theorems and the groupoid free energy density.
The rate distortion function $R(D)$ determines a channel capacity that is measurable in an analogous
way to a free energy that regulates many of the (inter)reactive /reciprocating processes that have been described.
Rate distortion arguments suggest that if an external information source is pathogenic, then sufficient exposure
to it within a developmental stage will likely result in a image inscribed on mind and body in a
punctuated fashion, subsequently causing a developmental dysfunction. In this analogy, the reduction of the $R(D)$ amounts
to `lower temperature' which in turn directs the system to behavioral patterns which are less enriched
and are less complex. Further, accurate and efficient communicating systems require a greater channel capacity,
and keeping in mind the analogy with free energy density, a higher rate of metabolism is necessary and further costs
are incurred (cf \S\ref{feynman}). Failure to provide such resources equates to a decline in processing, possibly
to a point of disintegration. On the other hand, increased communication between the system's cognitive modules,
depending on the availability of free energy, will usually be followed up by a phase transition (in essence, this
is what the holonomy groupoid encodes as shown in Glazebrook and Wallace (2009a)) inducing further complexity into the systems behavior.
In principle one might envisage an associated \emph{holonomy groupoid atlas} for interacting
cognitive-dynamical systems based on synchronous (geometric) phase transitions of the various constituents. This would amount to
a broad-scale descriptive artifact for the purpose of understanding the cumulative transitional mechanisms of living processes
that may eventually uncover some even deeper conceptual issues.

Stochastic processes are perhaps more in keeping with what the world expects compared to a strictly
deterministic approach, though the former are likely to entail
higher computational costs and some neuroscientific work in this area
is aimed at reducing such costs (cf Manninen et al., 2006).
Integration of the Onsager stochastic differential equation towards a multidimensional
It\^{o} process, leads naturally to a stochastic flow which in our formulation
diffuses across the atlas through mainly
noisy channels (cf \S\ref{multiphase}). Such evidence is provided by Tlusty's rate-distortion
analysis of the genetic coding map where it is shown that the evolution of the code unfolds as a
smooth flow on the codon space as realized in Tlusty (2007). With such examples in mind, we have
presented here a novel development, very much in tune with the ``flow'' nature of living
systems as once envisaged by Bertalanffy and Prigogine, while at the same time our approach
using rate distortion principles embraces
many central features of related theories of cognition such as those proposed
in Atlan and Cohen (1998), Baars (1988), Baars and Franklin (2003), Cohen and Harel (2007) and
Maturana and Varela (1980a, 1980b).

%%%%%%%%%%%%%%%%%%%%%%%%%%%%%%%%%%%%%%%%%%%%%%%%%%%%%%%%%%%%%%%%%%%%%%%%

\bigbreak

%\textbf{Acknowledgements.} We wish to thank the reviewers for
%their various comments and suggestions.

%%%%%%%%%%%%%%%%%%%%%%%%%%%%%%%%%%%%%%%%%%%%%%%%%%%%%%%%%%%%%%%%%%%%%%%%%

%\appendix

\section{Appendix: Probability space and Brownian motion}\label{appendix-A}

We state some basic details as to be found in Harrison (1985), Kunita (1990) and Protter (1985).
Let $\Omega$ be a set. A collection $\F$ of subsets of $\Omega$ is
called a \emph{$\sigma$-field} if it contains an empty set and it is closed under the operations of countable unions and
complements. It is customary to call $(\Omega, \F)$ a \emph{measurable space} in which members of
$\Omega$ are called \emph{samples} and those of $\F$ are called \emph{events}. Let $P$ be a $\sigma$-additive measure on
$(\Omega,\F)$. It is called a \emph{probability} if $P(\Omega)=1$. The triple $(\Omega, \F, P)$ is then called
a \emph{probability space}. Let $\bF=\{\F_t: t \geq 0\}$ be a family of $\sigma$-algebras on $\Omega$ such that~
a) $\F_t \subseteq \F$, for all $t \geq 0$, and b) $\F_s \subseteq \F_t$, if $s \leq t$. Then $\bF$ is said to be a
\emph{filtration} (an increasing sequence of sub-$\sigma$-algebras) of $(\Omega, \F)$. The filtration $\bF$ characterizes
how information arises (how uncertainty is resolved) and $\F_t$ may be interpreted as the set of all events whose occurrence or
nonoccurrence will be determined at time $t$. In a filtered probability space it is usually understood that
$\F = \F_{\infty}$.

A stochastic process $W = \{W_t\}, ~t \in \bT$, is said to be \emph{adapted} (relative to
$(\Omega, \bF, P)$) if $W_t$ is measurable with respect to $\F_t$, for all $t \geq 0$.
Loosely speaking, this means that the information available at time $t$ includes the history of $W$
up to that point. The \emph{predictable $\sigma$-field} is the least $\sigma$-field in the product space $[0,T] \times \Omega$
for which all continuous $F_t$-adapted processes are measurable. \emph{A predictable process} is then defined
as a process that is measurable with respect to the predictable $\sigma$-field. An example is a
continuous $\F_t$-adapted process.
Recall that a process of random variables $W = \{W_t\}, t \in \bT$ is Brownian motion (viz a Weiner process) if and only if
\begin{itemize}
\item[(1)] $W_0= 0$ with probability $1$.

\item[(2)] For $0 \leq s < t < \infty$ the increment $W_t - W_s$ is normally distributed $N(0,\vert t-s \vert)$.

\item[(3)] For $0 \leq t_0 < t_1 < \cdots < t_n < \infty$, the set of increments
$$
\{ W_{t_0}, W_{t_j} - W_{t_j -1}, \text{for}~ 1 \leq j \leq k \},
$$
is a set of independent random variables (that is, the increments are independent of the past).
\end{itemize}
A process $Q$ is called a \emph{$(\mu, \sigma)$-Brownian motion} if it has the form $Q_t = Q_0 + \mu t + \sigma W_t$
where $W$ is a Weiner process and $Q_0$ is independent of $W$. Then we have $Q_{t+s} - Q_t \sim N(\mu s, \sigma^2 s)$.

\section{Appendix: Basic results of information theory}\label{appendix-B}

\subsection{The Shannon uncertainties}\label{uncertainty}

Invoking the spirit of the Shannon-McMillan Theorem, it is
possible to define an APSE information source $\mathbf{X}$
associated with stochastic variates $X_{j}$ having joint and
conditional probabilities $P(a_{0}, \ldots,a_{n})$ and
$P(a_{n}|a_{0},\ldots,a_{n-1})$ such that appropriate joint and
conditional Shannon uncertainties satisfy the classic relations
\begin{equation}
\begin{aligned}
H[\mathbf{X}] &=\lim_{n \lra \infty} \frac{\log[N(n)]}{n}
\\
&= \lim_{n \lra \infty}H(X_{n}|X_{0}, \ldots,X_{n-1})\\ &= \lim_{n
\lra \infty} \frac{H(X_{0},\ldots ,X_{n})}{n}.
\end{aligned}
\end{equation}
This information source is defined as \textit{dual} to the
underlying ergodic cognitive process (Wallace, 2005).

Recall that the Shannon uncertainties $H(\dots)$ are
cross-sectional law-of-large-numbers sums of the form $-\sum_{k}
P_{k}\log[P_{k}]$, where the $P_{k}$ constitute a probability
distribution (for the basic details, see Ash,1990; Berger, 1971; Cover and Thomas, 1991; Khinchin, 1957).
Messages from an information source, seen as symbols $x_{j}$ from
some alphabet, each having probabilities $P_{j}$ associated with a
random variable $X$, are `encoded' into the language of a
`transmission channel', a random variable $Y$ with symbols
$y_{k}$, having probabilities $P_{k}$, possibly with error.
Someone receiving the symbol $y_{k}$ then retranslates it (without
error) into some $x_{k}$, which may or may not be the same as the
$x_{j}$ that was sent. More formally, the message sent along the
channel is characterized by a random variable $X$ having the
distribution
\begin{equation}\label{distrib1}
P(X=x_{j})=P_{j}, j=1, \ldots,M.
\end{equation}
The channel through which the message is sent is characterized by
a second random variable $Y$ having the distribution
\begin{equation}\label{distrib2}
P(Y=y_{k})=P_{k}, k=1, \ldots,L.
\end{equation}
Let the joint probability distribution of $X$ and $Y$ be defined
as
\begin{equation}
P(X=x_{j},Y=y_{k})=P(x_{j},y_{k})=P_{jk},
\end{equation}
and the conditional probability of $Y$ given $X$ as
\begin{equation}
P(Y=y_{k}\vert X=x_{j}) = P(y_{k}\vert x_{j}).
\end{equation}
Then the Shannon uncertainty of $X$ and $Y$ independently and the
joint uncertainty of $X$ and $Y$ together are defined respectively
as
\begin{equation}\label{info20}
\begin{aligned}
H(X) &=-\sum_{j=1}^M P_{j}\log(P_{j}), \\ H(Y) &=-\sum_{k=1}^L
P_{k}\log(P_{k}), \\ H(X,Y) &=-\sum_{j=1}^M \sum_{k=1}^L P_{j,k}
\log(P_{jk}).
\end{aligned}
\end{equation}
The \textit{conditional uncertainty} of $Y$ given $X$ is defined
as
\begin{equation}\label{info21}
H(Y|X)=-\sum_{j=1}^M \sum_{k=1}^L P_{jk} \log[P(y_{k}|x_{j})].
\end{equation}
For any two stochastic variates $X$ and $Y$, we have the
inequality $H(Y) \geq H(Y|X)$, as the knowledge of $X$ generally
gives some knowledge of $Y$. Equality occurs only in the case of
stochastic independence. Since $P(x_{j},
y_{k})=P(x_{j})P(y_{k}\vert x_{j})$, it is deduced
\begin{equation}
H(X|Y)=H(X,Y) - H(Y).
\end{equation}
The information transmitted by translating the variable $X$ into
the channel transmission variable $Y$ -- possibly with error --
and then retranslating without error the transmitted $Y$ back into
$X$, is defined as
\begin{equation}\label{info22}
\begin{aligned}
I(X \vert Y) &\equiv H(X) - H(X \vert Y)\\
 &= H(X) + H(Y) - H(X,Y),
\end{aligned}
\end{equation}
where we refer to Berger (1971), Cover and Thomas (1991) and Khinchin (1957) for details.  The
essential point is that if there is no uncertainty in $X$ given
the channel $Y$, then there is no loss of information through
transmission. In general this will not be true, and herein lies
the essence of the theory.

\subsection{The Rate Distortion Theorem}\label{RDTheorem}

Following Wallace (2005), suppose we have an (ergodic)
information source $Y$ with output from a particular alphabet
generating sequences of the form
\begin{equation}
y^n = y_1, \ldots, y_n
\end{equation}
`digitalized' in some sense, and induce a chain of `digitalized'
values
\begin{equation}
b^n = b_1, \ldots, b_n
\end{equation}
where the $b$--alphabet is considered more restricted than the
$y$--alphabet. In this way, $b^n$ is deterministically
retranslated into a reproduction of the signal $y^n$~. That is,
each $b^n$ is mapped onto a unique $n$--length $y$--sequence in
the alphabet of $Y$:
\begin{equation}
b^m \lra \hat{y}^n = \hat{y}_1, \ldots, \hat{y}_n.
\end{equation}
We remark that many $y^n$ sequences may be mapped onto the same
retranslation sequence $\hat{y}^n$, the set of which is denoted
$\what{Y}$; this may be interpreted as a loss of information.

A distortion measure $d : Y \times \what{Y} \lra \mathbb R^{+}$,
between paths $y^n$ and $\hat{y}^n$ is defined as
\begin{equation}\label{distmeasure}
d (y^n, \hat{y}^n) = \frac{1}{n} \sum^n_{i=1} \ul{d}(y_j,
\hat{y}_j),
\end{equation}
for some suitable distance function $\ul{d}$ (such as the Hamming
distance). Suppose that with each path $y^n \in Y$ and each
$b^n$--path retranslation $\hat{y}^n \in \what{Y}$ into the
$y$--language, we consider the associated individual, joint, and
conditional probability distributions
\begin{equation}
p(y^n)~,~ p(\hat{y}^n)~,~ p(y^n \vert \hat{y}^n).
\end{equation}
The average distortion is then defined to be
\begin{equation}\label{avdistort}
D = \sum_{y^n} p(y^n)~d (y^n, \hat{y}^n).
\end{equation}
For the corresponding strings $Y$ (incoming), $\what{Y}$
(outgoing), applying the Shannon uncertainty rule of
\eqref{info22} gives
\begin{equation}
\begin{aligned}
I(Y, \what{Y}) &\equiv H(Y) - H(Y \vert \what{Y})\\
 &= H(Y) +
H(\what{Y}) - H(Y, \what{Y}).
\end{aligned}
\end{equation}
The \emph{information rate distortion function} $R(D)$ for a
source sequence $Y$, retranslated sequence $\what{Y}$ with
distortion measure $d : Y \times \what{Y} \lra \mathbb{R}^{+}$, is
defined as follows.
Let $\Upsilon = \sum_{(y, \hat{y})}~ p(y)~ p(y \vert \hat{y})~ d
(y, \hat{y})$. Then
\begin{equation}\label{rdfunction}
R(D) = \sum_{p(y, \hat{y})~ :~ \Upsilon \leq D} I(Y, \what{Y}).
\end{equation}
To explain this notation, the minimization is over all conditional
distributions $p(y \vert \hat{y})$, for which the joint
distribution $p(y, \hat{y}) = p(y)~ p(y \vert \hat{y})$ satisfies
average distortion less than or equal to $D$.
The {Rate Distortion Theorem} (see e.g. Berger, 1971; Cover and Thomas, 1991) states that \emph{$R(D)$ is the minimum
necessary rate of information transmission (effectively the
channel capacity) so that the average distortion does not exceed
the distortion $D$}.

%%%%%%%%%%%%%%%%%%%%%%%%%%%%%%%%%%%%%%%%%%%%%%%%%%%%%%%%%%%%%%%%%%%%%%%%%%%%

\section{References}
\label{refs}
%\begin{thebibliography}{99}

%\bibitem{Ash}
\medn
Ash, R., 1990, \textit{Information Theory}, Dover Publications,
New York.

\medn
%\bibitem{Atlan}
Atlan, H., and Cohen, I., 1998, Immune information,
self-organization and meaning, \textit{International Immunology},
\textbf{10}, 711-717.

\medn
%\bibitem{Baars1}
Baars, B., 1988, \textit{A Cognitive Theory of Consciousness},
Cambridge University Press, New York.

\medn
%\bibitem{Baars2}
Baars, B. and Franklin, S., 2003, How conscious experience and working memory
interact, \textit{Trends in Cognitive Science} \textbf{7}(4), 166--172.

\medn
%\bibitem{BBGG}
Baianu, I. C., Brown, R., Georgescu, G., and Glazebrook, J. F., 2006,
Complex nonlinear biodynamics in categories, higher dimensional
algebra and Lukasiewicz--Moisil topos: transformations of
neuronal, genetic and neoplastic networks, \textit{Axiomathes},
\textbf{16} Nos. 1--2, 65--122.

\medn
%\bibitem{Bak}
Bak, A., Brown, R., Minian, G., and Porter, T., 2006, Global
actions, groupoid atlases and related topics, \textit{Journal of
Homotopy and Related Structures}, \textbf{1}, 1-54.

\medn
%\bibitem{Bennett}
Bennett, C. H., 1982, The thermodynamics of computation: a Review,
\textit{International Journal of Theoretical Physics,}
\textbf{21} (12), 905--940.

\medn
%\bibitem{Berger}
Berger, T., 1971, \textit{Rate Distortion Theory: A mathematical basis for data compression},
Prentice--Hall, Inc., Englewood Cliffs, NJ.

%%\bibitem{BGH}
%%Bolker E. D., Guillemin, V. W., and Holm, T. S., 2006, How is a
%%graph like a manifold?, to appear.
%%\\http://arxiv:math.CO/0206103

\medn
%\bibitem{Bertal}
Bertalanffy, L., 1972, \textit{General System Theory: foundations, development, applications}, George
Braziller, New York.

\medn
%\bibitem{Brown1}
Brown, R., 2006, \emph{Topology and Groupoids}, BookSurge LLC.

\medn
%\bibitem{Burago}
Burago, D., Burago, Y., and Ivanov, S., 2001, \textit{A Course in
Metric Geometry}, American Mathematical Society, Providence, RI.

\medn
%\bibitem{Cao}
Cao, Y. U., Fukunaga, A. S., Kahng, A. B., and Meng, F., 1997,
Cooperative mobile robotics: Antecedents and directions,
\textit{Autonom. Robots}, \textbf{4}, 7--27.

\medn
%\bibitem{Capra}
Capra, F. 1996, \textit{The Web of Life: A new scientific understanding of living systems},
Anchor Books, New York.

\medn
%\bibitem{Caspi1}
Caspi, A. and Moffit, T., 2006, Gene-environment interactions in psychiatry:
joining forces with neuroscience, \textit{Nature Reviews Neuroscience} \textbf{7}, 583-590.

\medn
%\bibitem{Clark}
Clark, A., 1997, \textit{Being There: Putting Brain, Body and
World Together Again}, MIT Press, Cambridge, MA.

\medn
%\bibitem{Cohen1}
Cohen, I., 2000, \textit{Tending Adam's Garden: Evolving the
Cognitive Immune Self}, Academic Press, New York.

\medn
%\bibitem{Cohen2}
Cohen, I., and Harel D., 2007, Explaining a complex living system:
dynamics, multiscaling and emergence, \textit{Journal of The Royal Society
Interface} \textbf{4}, 175--182.

\medn
%\bibitem{ConnesBook}
Connes, A., 1994, \textit{Noncommutative Geometry}, Academic Press,
San Diego, CA.

\medn
%\bibitem{Cover}
Cover, T., and Thomas, J., 1991, \textit{Elements of Information
Theory}, John Wiley and Sons, New York.

\medn
%\bibitem{Debord}
Debord, C., 2001, Holonomy groupoids of singular foliations, \textit{J. Differential Geom.},
\textbf{58}(3), 467--500.

\medn
%\bibitem{Dretske1}
Dretske, F., 1981, \textit{Knowledge and the Flow of Information},
MIT Press, Cambridge, MA.

\medn
%\bibitem{Dretske2}
Dretske, F., 1988, \textit{Explaining Behavior}, MIT Press,
Cambridge, MA.

\medn
%\bibitem{Durham}
Durham, W., 1991, \emph{Coevolution: Genes, Culture and Human Diversity}, Stanford Univ. Press, Palo Alto, CA.

\medn
%\bibitem{Feynman}
Feynman, R., 1996, \textit{Feynman Lectures on Computation},
Addison-Wesley, Reading, MA.

\medn
%\bibitem{Gamma}
Gammaitoni, L., H\"anggi, P., Jung, P., and Marchesoni, F., 1998,
Stochastic resonance, \textit{Rev. Mod. Phys.} \textbf{70}(1),
223--287.

\medn
%\bibitem{Gilbert1}
Gilbert, P., 2006, Evolution and depression: issues and implications,
\textit{Psychological Medicine}, \textbf{36}, 287--297.

\medn
%\bibitem{Glade}
Glade, N., Ben Amor, H. M., and Bastien, O., 2009,
Trail systems as fault tolerant wires and their use in bio--processors.
In Amar, P. et al., (Eds.) \textit{Modeling Complex Biological Sytems in the Context of Genomics,
Proceedings of the Spring School, Nice 2009}, 85--119.

\medn
%\bibitem{GW1}
Glazebrook, J. F., and Wallace, R., 2009a, Small worlds
and red queens in the global
workspace: an information--theoretic approach,
\textit{Cognitive Systems Research}, \textbf{10},
333--365.

\medn
%\bibitem{GW2}
Glazebrook, J. F., and Wallace, R., 2009b, Rate distortion
manifolds as model spaces for cognitive information, \textit{Informatica} \textbf{33} (2009), 309--345.

\medn
%\bibitem{Golubitsky}
Golubitsky, M., and Stewart, I., 2006, Nonlinear dynamics and
networks:the groupoid formalism, \textit{Bulletin of the American
Mathematical Society}, \textbf{43}, 305-364.

\medn
%\bibitem{Harrison}
Harrison, J. M., 1985, \textit{Brownian Motion and Stochastic Flow Systems}, J. Wiley, New York.

\medn
%\bibitem{Hollan}
Hollan, J., Hutchins, J., and Kirsch, D., 2000, Distributed
cognition: toward a new foundation for human-computer interaction
research, \textit{ACM Transactions on Computer-Human Interaction},
\textbf{7}, 174-196.

\medn
%\bibitem{Hoyo}
del Hoyo, M. L., and Minian, E. G., 2009, Classical invariants for global actions and
groupoid atlases, \textit{Appl. Categorical Structures}, \textbf{16} (6), 689--721.

\medn
%\bibitem{Hutchins}
Hutchins, E., 1994, \textit{Cognition in the Wild}, MIT Press,
Cambridge, MA.

\medn
%\bibitem{Jablonka}
Jablonka, E. and Lamb, M., 2005, \textit{Evolution in Four Dimensions},
MIT Press, Cambridge, MA.

\medn
%\bibitem{Jerne}
Jerne, N. K., 1974, Towards a network theory of the immune system,
\emph{Ann. Immunology (Inst. Pasteur)}, {125C}, 373--389.

\medn
%\bibitem{Khinchin}
Khinchin, A., 1957, \textit{The Mathematical Foundations of
Information Theory}, Dover Publications, New York.

\medn
%\bibitem{Krieger}
Krieger, M. J. B., Billeter. J.-B., and Keller, L., 2000, Ant--like task allocation and recruitment
in cooperative robots, \textit{Science}, \textbf{406}, 992--995.

\medn
%\bibitem{Kunita}
Kunita, H., 1990, \textit{Stochastic flows and stochastic differential equations},
Cambridge Studies in Advanced Mathematics \textbf{24}, Cambridge University Press.

\medn
%\bibitem{Kurz}
Kurzynski, M., 2006, \textit{The Thermodynamic Machinery of Life},
Springer--Verlag, Berlin Heidelberg New York.

\medn
%\bibitem{Laing}
Laing, R. D., and Esterson, A., 1964, \emph{Sanity, Madness and the Family}, Tavistock Press, London.

\medn
%\bibitem{Landau}
Landau, L., and Lifshitz, E., 2007, \textit{Statistical Physics} (I) (3rd Ed.),
Elsevier, New York.

\medn
%\bibitem{Laughlin}
Laughlin, C. D. Jr., and d'Aquili, E. G., 1974, \textit{Biogentic Structuralism},
Columbia University Press.

\medn
%\bibitem{Li}
Li, M., and Vit\'{a}nyi, P., 1993, \emph{An Introduction to Kolmogorov
Complexity and its Applications},
Texts and Monographs in Computer Science, Springer-Verlag, New York.

\medn
%\bibitem{Li-Vit}
Li, M., and Vit\'{a}nyi, P. 1992, {The theory of thermodynamics of computation}, in
\emph{Proc. IEEE Physics of Computation Workshop}, IEEE Publ., 42--46.

%\bibitem{Mack}
%Mack A., 1998, \textit{Inattentional Blindness}, (Cambridge, MA,
%MIT Press).

\medn
%\bibitem{Manninen}
Manninen, T., Linne, M-J, and Ruohonen, K., 2006, Developing It\^{o} stochastic differential equation models for neuronal signal transduction in pathways, \textit{Comp. Biology and Chem.} \textbf{30}, 280--291.

\medn
%\bibitem{Margulis}
Margulis, L., 2004, Serial endosymbiotic theory (SET) and composite individuality.
Transition from bacterial to eukaryotic genomes,
\textit{Microbiology Today}, \textbf{31}, 173-174.

\medn
%\bibitem{MV1}
Maturana, H. R., and Varela, F. J., 1980a, \emph{Autopoiesis and
Cognition--The Realization of the Living}, Boston Studies in the
Philosophy of Science Vol. 42, Reidel Pub. Co. Dordrecht.

\medn
%\bibitem{MV2}
Maturana, H. R., and Varela, F. J., 1980b, \emph{The Tree of Knowledge},
Shambhala Publications, Boston, MA.

\medn
%\bibitem{Mingers}
Mingers, J., 1991, The cognitive theories of Maturana and Varela,
\textit{Systems Practice}, \textbf{4}(4), 319--338.

\medn
%\bibitem{MMr}
Moerdijk, I., and Mr\v{c}un, J., 2003, \emph{Introduction to
Foliations and Lie Groupoids}, Cambridge Studies in Adv. Math. {91},
Cambridge Univ. Press.

\medn
%\bibitem{Moffitt}
Moffitt, T. E., Caspi, A., and Rutter, M., 2006, Measured gene-environment interactions in
psychopathology: Concepts, research strategies, and implications for research, intervention, and
public understanding of genetics, \textit{Perspectives on Psychological Science}, \textbf{1}(1),
5--27.

\medn
%\bibitem{Nisbett}
Nisbett, R., 2003, \textit{The Geography of Thought: How Asians and Westerners
think differently ... and why}, Free Press, New York.

\medn
%\bibitem{Prigogine}
Prigogine, I., 1980, \emph{From Being to Becoming: Time and
Complexity in the Physical Sciences}, W. H. Freeman and Co.,  San
Francisco.

\medn
%\bibitem{Protter}
Protter, P., 1995, \textit{Stochastic Integration and Differential Equations:
A New Approach}, Springer, New York.

\medn
%\bibitem{Richerson}
Richerson, P., and Boyd, R., 2004, \textit{Not by Genes Alone: How
Culture Transformed Human Evolution}, Chicago University Press,
Chicago.

\medn
%\bibitem{Stewart1}
Stewart, I., Golubitsky, M., and Pivato, M., 2003, Symmetry
groupoids and patterns of synchrony in coupled cell networks,
\textit{SIAM Journal of Applied Dynamical Systems}, \textbf{2},
609-646.

\medn
%\bibitem{Tlusty1}
Tlusty, T., 2007, A model for the emergence of the genetic code as a transition
in a noisy channel, \textit{Journal of Theoretical Biology} \textbf{249}, 331--342.

\medn
%\bibitem{Tlusty2}
Tlusty, T., 2008, Rate-distortion scenario for the emergence and evolution
of noisy molecular codes, \textit{Physical Review Letters} \textbf{100},
048101 (1-4).

\medn
%\bibitem{Wallace3}
Wallace, R., 2005, \textit{Consciousness: A Mathematical Treatment
of the Global Neuronal Workspace Model}, Springer, New York.

\medn
%\bibitem{Wallace4}
Wallace, R., 2005, A global workspace perspective on mental
disorders, \textit{Theoretical Biology and Medical Modelling},
\textbf{2}, 49.

\medn
%\bibitem{Wallace7}
Wallace, R., 2007, Culture and inattentional blindness: a global
workspace perspective, \textit{Journal of Theoretical Biology},
\textbf{245}, 378-390.

\medn
%\bibitem{Wallace8}
Wallace, R., 2009a, The cultural epigenetics of
psychopathology: The missing heritability of complex diseases found?,
to appear.

\medn
%\bibitem{Wallace9}
Wallace, R., 2009b, Roman roads: The hierarchial endosymbiosis of
cognitive modules, to appear.

%\bibitem{Wallaces0}
%Wallace, D., and Wallace, R., 1998, \emph{A Plague on Your Houses},
%Verso Press, New York.

\medn
%\bibitem{Wallaces1}
Wallace, R., and Wallace R.G., 1998, Information theory,
scaling laws, and the thermodynamics of evolution, \textit{Journal
of Theoretical Biology}, \textbf{192}, 545-559.

\medn
%\bibitem{Wallaces2}
Wallace, R., and Wallace R.G., 1999, Organisms, organizations
and interactions: an information theory approach to biocultural
evolution, \textit{BioSystems}, \textbf{51}, 101-119.

\medn
%\bibitem{Wallaces3}
Wallace, R., and Wallace, D., 2009, \textit{Gene Expression and its Discontents:
The social production of pandemic chronic disease}, Springer, New York.

\medn
%\bibitem{Wallaces4}
Wallace, R., and Wallace, D., 2008, Punctuated equilibrium in statistical models of
generalized coevolutionary resilience: how sudden ecosystem transitions can entrain both phennotype expression
and Darwinian selection,
\textit{Transactions on Computational Systems Biology IX}, LNBI 5121: 23 --85.

\medn
%\bibitem{WallaceFullilove}
Wallace, R., and Fullilove M., 2008, \emph{Collective
Consciousness and Its Discontents: Institutional distributed
cognition, racial policy, and public health in the United States},
Springer, New York.

\medn
%\bibitem{Wang}
Wang, M., and Uhlenbeck, G., 1945, On the theory of the Brownian motion II,
\textit{Reviews of Modern Physics}, \textbf{17}, 323--342.

\medn
%\bibitem{Weinstein}
Weinstein, A., 1996, Groupoids: unifying internal and external
symmetry, \textit{Notices of the American Mathematical Society},
\textbf{43}, 744-752.

\medn
%\bibitem{West}
West, R. L., Stewart T. C., Lebiere C., and Chandrasekharan, S.
(2005), Stochastic resonance in human cognition: ACT-R versus game
theory, associative neural networks, recursive neural networks,
Q-learning and humans. In Bara, B., Barsalou, L., \& Bucciarelli, M.
(Eds.), \textit{Proceedings of the 27th Annual Conference of the
Cognitive Science Society-Stresa, Italy, 2005} (pp. 2353-2358).
Mahwah, NJ: Lawrence Erlbaum Associates.

\medn
%\bibitem{Witzany}
Witzany, G., 2006, Serial endosymbiotic theory (SET): the biosemiotic update, \textit{Acta Biotheoretica},
\textbf{54}, 103--117.

%\end{thebibliography}

\end{document}